\def\llncs{0}
\def\showtableofcontents{1}
\documentclass[11pt]{article}
\usepackage{graphicx} %
\usepackage{fullpage}
\usepackage{amsmath}

\usepackage{amsthm}
\usepackage{aliascnt}
\usepackage{amssymb}
\usepackage{physics}
\usepackage[colorlinks=true,linkcolor=blue,urlcolor=red,citecolor=blue]{hyperref}
\usepackage[capitalise,nameinlink]{cleveref}
\usepackage{dsfont}
\usepackage{xcolor}
\usepackage{mathtools}
\usepackage{breakcites}
\usepackage{subcaption}
\usepackage[shortlabels]{enumitem}

\ifnum\llncs=0
\newtheorem{theorem}{Theorem}[section]
\newaliascnt{proper}{theorem}

\aliascntresetthe{proper}
\crefname{proper}{Property}{Properties}
\Crefname{proper}{Property}{Properties}

\newaliascnt{corollary}{theorem}
\newtheorem{corollary}[corollary]{Corollary}
\aliascntresetthe{corollary}
\crefname{corollary}{Corollary}{Corollaries}
\Crefname{corollary}{Corollary}{Corollaries}

\newaliascnt{lemma}{theorem}
\newtheorem{lemma}[lemma]{Lemma}
\aliascntresetthe{lemma}
\crefname{lemma}{Lemma}{Lemmas}
\Crefname{lemma}{Lemma}{Lemmas}

\newaliascnt{conjecture}{theorem}
\newtheorem{conjecture}[conjecture]{Conjecture}
\aliascntresetthe{conjecture}
\crefname{conjecture}{Conjecture}{Conjectures}
\Crefname{conjecture}{Conjecture}{Conjectures}

\newaliascnt{claim}{theorem}
\newtheorem{claim}[claim]{Claim}
\aliascntresetthe{claim}
\crefname{claim}{Claim}{Claims}
\Crefname{claim}{Claim}{Claims}

\newaliascnt{proposition}{theorem}

\aliascntresetthe{proposition}
\crefname{proposition}{Proposition}{Propositions}
\Crefname{proposition}{Proposition}{Propositions}

\theoremstyle{definition}

\newaliascnt{definition}{theorem}
\newtheorem{definition}[definition]{Definition}
\aliascntresetthe{definition}
\crefname{definition}{Definition}{Definitions}
\Crefname{definition}{Definition}{Definitions}

\newaliascnt{remark}{theorem}
\newtheorem{remark}[remark]{Remark}
\aliascntresetthe{remark}
\crefname{remark}{Remark}{Remarks}
\Crefname{remark}{Remark}{Remarks}

\newaliascnt{note}{theorem}

\aliascntresetthe{note}
\crefname{note}{Note}{Notes}
\Crefname{note}{Note}{Notes}

\newaliascnt{notation}{theorem}

\aliascntresetthe{notation}
\crefname{notation}{Notation}{Notations}
\Crefname{notation}{Notation}{Notations}

\newaliascnt{observation}{theorem}

\aliascntresetthe{observation}
\crefname{observation}{Observation}{Observations}
\Crefname{observation}{Observation}{Observations}

\newaliascnt{invariant}{theorem}

\aliascntresetthe{invariant}
\crefname{invariant}{Invariant}{Invariants}
\Crefname{invariant}{Invariant}{Invariants}

\newaliascnt{construction}{theorem}
\newtheorem{construction}[construction]{Construction}
\aliascntresetthe{construction}
\crefname{construction}{Construction}{Constructions}
\Crefname{construction}{Construction}{Constructions}
\fi

\newcommand{\A}{\mathcal{A}}

\newcommand{\D}{\mathcal{D}}
\renewcommand{\O}{\mathcal{O}}
\newcommand{\N}{\mathbb{N}}

\newcommand{\E}{\mathop{\mathbb{E}}}

\newcommand{\U}{\mathcal{U}}

\newcommand{\OracleA}{\mathsf{A}}
\newcommand{\OracleB}{\mathsf{B}}
\newcommand{\OracleC}{\mathsf{C}}

\newcommand{\F}{\mathbb{F}}

\newcommand{\poly}{\mathsf{poly}}
\newcommand{\negl}{\mathsf{negl}}
\newcommand{\polylog}{\mathsf{polylog}}
\newcommand{\quasipoly}{\mathsf{quasipoly}}

\newcommand{\wt}[1]{\widetilde{#1}}
\newcommand{\ol}[1]{\overline{#1}}
\newcommand{\floor}[1]{\left\lfloor #1 \right\rfloor}
\newcommand{\ceil}[1]{\left\lceil #1 \right\rceil}

\DeclareMathOperator{\Dom}{\mathsf{Dom}}
\DeclareMathOperator{\Sim}{\mathsf{Sim}}

\DeclareMathOperator{\Dec}{\mathsf{Dec}}

\DeclareMathOperator{\Ver}{\mathsf{Ver}}
\DeclareMathOperator{\SG}{\mathsf{SG}}
\DeclareMathOperator{\Expand}{\mathsf{Expand}}
\DeclareMathOperator{\Mix}{\mathsf{Mix}}
\DeclareMathOperator{\Init}{\mathsf{Init}}
\DeclareMathOperator{\Test}{\mathsf{Test}}
\DeclareMathOperator{\Samp}{\mathsf{Samp}}
\DeclareMathOperator{\OPT}{\mathsf{OPT}}
\DeclareMathOperator{\Eval}{\mathsf{Eval}}
\DeclareMathOperator{\ML}{\mathsf{ML}}
\DeclareMathOperator{\MAJ}{\mathsf{MAJ}}

\newcommand{\xor}{\oplus}

\newcommand{\Haar}{\mathsf{Haar}}
\newcommand{\HaarSt}{\mathsf{HaarSt}}

\newcommand{\Reg}[1]{\mathbf{#1}}

\newcommand{\Hyb}{\mathsf{Hyb}}
\newcommand{\aux}{\mathsf{aux}}

\renewcommand{\P}{\mathsf{P}}
\newcommand{\NP}{\mathsf{NP}}
\newcommand{\BQP}{\mathsf{BQP}}
\newcommand{\BPP}{\mathsf{BPP}}
\newcommand{\PSPACE}{\mathsf{PSPACE}}
\newcommand{\PP}{\mathsf{PP}}
\newcommand{\QCMA}{\mathsf{QCMA}}
\newcommand{\IP}{\mathsf{IP}}
\newcommand{\AC}{\mathsf{AC}}
\newcommand{\PH}{\mathsf{PH}}

\newcommand{\RO}{\mathrm{RO}}

\newcommand{\Hilb}[1]{\mathcal{H}(\{0,1\}^{#1})}

\title{Quantum Advantage has Few Complexity Implications}
\author{
    Eli Goldin\\
    {\small Princeton University}
    \and
    Alex Lombardi\\
    {\small Princeton University}
    \and
    Tomoyuki Morimae\\
    {\small Yukawa Institute for Theoretical Physics}\\
    {\small Kyoto University}
}

\date{November 3, 2026}
\begin{document}

\hypersetup{pageanchor=false}
\maketitle
\thispagestyle{empty}

\begin{abstract}
    We consider the complexity implications of proofs of quantumness and (quantum verifier) oblivious state preparation. We show that both of these objects can coexist with the easiness of nondeterministic computation. That is, we give evidence that proofs of quantumness can exist when $\P = \NP$ and that (quantum verifier) oblivious state preparation can exist when $\BQP = \QCMA$.

    In particular, we give a classical oracle relative to which
    \begin{enumerate}
        \item $\P^\O = \NP^\O$
        \item $\BQP^\O = \PSPACE^\O$
        \item Proofs of quantumness exist. In particular, there exists a language in $\IP^\O \subseteq \BQP^\O$ which is hard on average against $\P^\O/poly$.
    \end{enumerate}

    We give a unitary oracle such that
    \begin{enumerate}
        \item $\BQP^\O = \QCMA^\O$
        \item Quantum verifier oblivious state preparation exists.
    \end{enumerate}

    These oracle separations directly imply a number of related consequences. For example, our first result implies that proofs of quantumness cannot be used to construct classical or quantum cryptography with a relativizing proof. Our second result implies that quantum computation/classical communication oblivious transfer cannot be used to construct post-quantum one-way functions with a relativizing proof.
\end{abstract}

\ifnum\showtableofcontents=1
{
\newpage
\setcounter{tocdepth}{2}
\tableofcontents
\thispagestyle{empty}
}
\newpage
\fi

\setcounter{page}{1}
\hypersetup{pageanchor=true}
\section{Introduction}

Quantum advantage, the ability of a quantum computer to perform a task that no efficient classical computer can, is the defining promise of quantum computation. For this promise to be checkable, the task should admit efficient classical verification. Sampling output distributions of quantum computers is believed to be classically hard~\cite{STOC:AarArk11,BJS11}, but it is not efficiently verifiable. In contrast, a \emph{proof of quantumness} (PoQ)~\cite{FOCS:BCMVV18} is an interactive protocol between a probabilistic polynomial-time (PPT) verifier and a prover such that an honest quantum polynomial-time (QPT) prover makes the verifier accept with high probability, while no PPT prover can. 

In this paper, we study PoQ as well as both stronger and weaker forms of quantum advantage. On the stronger side, we consider the notion of oblivious state preparation (OSP)~\cite{C:BarKhu25}. OSP is a two-party protocol with classical communication in which a classical verifier can remotely generate a quantum state in the prover's register in such a way that only the verifier knows its classical description. This can be seen as a very strong form of quantum advantage: a prover who can obliviously prepare arbitrary states can demonstrate quantum advantage by giving the result of a related quantum measurement. In particular, this is a form of advantage which retains some notion of hardness even against quantum adversaries.

Beyond the fact that OSP can be used to construct PoQ~\cite{C:BarKhu25}, it has a number of other powerful applications. For example, OSP can be used to construct (single-qubit) proofs of quantum memory (PoQM)~\cite{C:BarKhu25,EPRINT:HMOY25}. A proof of quantum memory is a proof of quantumness where not only is a classical prover unable to make the verifier accept, but even a quantum prover with insufficient quantum memory cannot succeed. This means that an honest prover demonstrates the ability to maintain quantum memory of a specified size for a specified period during the protocol. In addition, OSP can be used to construct a powerful quantum cryptographic object known as search-secure quantum-computation-classical-communication (QCCC) oblivious transfer~\cite{C:BarKhu25}.

On the weaker side, we consider quantum-verifier proofs of quantumness (qv-PoQ) and quantum-verifier oblivious state preparation (qv-OSP), where the honest verifier in the protocol also has quantum capabilities. Note that these are particularly interesting in the latter setting, where the adversaries are also allowed to be quantum. In particular, as the corresponding protocols allow for quantum verification, qv-OSP is sufficient to instantiate the constructions of (quantumly verifiable) PoQM as well as search-secure QCCC oblivious transfer.

\paragraph{Constructions of PoQ and OSP.}
Since unconditionally sound proofs of quantumness are not known to exist (indeed, there is a complexity theoretic barrier, see below), PoQ are constructed and proved sound under computational hardness assumptions; this means that the quantum advantage is ``lost'' if the assumptions turn out to be false. The simplest PoQs arise from problems in $\NP \cap \BQP$ that are believed to be classically hard on average. The verifier samples an instance of the problem and the prover produces a witness. Examples of this formula include factoring and discrete log~\cite{FOCS:Shor94}. Proofs of quantumness can also be built from cryptographic assumptions such as the learning with errors (LWE) assumption~\cite{FOCS:BCMVV18}, quantum fully homomorphic encryption~\cite{STOC:KLVY23}, and indistinguishability obfuscation (iO) plus one-way functions (OWFs)~\cite{ITCS:MorYam23,C:ShmZha25}. Post-quantum versions of the above listed assumptions can all also be used to achieve OSP through similar constructions~\cite{C:BarKhu25,C:ShmZha25,ITCS:MorYam23,CWX26}. Unlike the $\NP\cap \BQP$ style of construction, these protocols are inherently interactive.

As far as the authors are aware, all known candidate constructions of PoQ rely on assumptions which themselves imply that classically-secure one-way functions exist, and so $ \NP \not\subset \BPP$. OSP itself directly implies that $\BQP \neq \QCMA$ (see~\Cref{sec:commitment-ev-puzzles}) in the classical verifier setting. This implication is not known to hold for qv-OSP.

\paragraph{Known complexity-theoretic/cryptographic lower bounds}
Despite the fact that all constructions of PoQ imply $\NP \not\subset \BPP$, it is open whether such an implication holds for PoQ generically. Likewise, it is not known whether the existence of qv-OSP implies $\BQP \neq \QCMA$. Nevertheless, complexity theoretic lower bounds are known for both primitives. It is known that if $\BPP = \PP$, then PoQ cannot exist~\cite{STOC:MorShiYam25,EPRINT:GMMY24}. Intuitively, this is because a classical prover with access to a $\PP$ oracle can simulate arbitrary quantum computation, even conditioned on partial outputs. Thus, a PPT algorithm can simulate the honest quantum prover. Similarly, if $\BQP = \PP$, then qv-OSP cannot exist~\cite{C:BarKhu25,EPRINT:HMOY25,EPRINT:GMMY24}.

We remark that for PoQ, stronger complexity lower bounds are known in restricted settings. If $\BPP=\NP$, then \textit{constant-round} PoQ cannot exist\footnote{We assumed this was a known result, but were unable to find a citation. As such, we have included a proof in \cref{coro:no_const_PoQ}.}. Furthermore, if one assumes a particular structure on the proof of security of the PoQ (namely that it is a black-box reduction to a quantumly-secure falsifiable assumption), then it is necessary that $\BQP \neq \PP$~\cite{TZ26,EPRINT:GMMY24}.

\paragraph{Conjectural lower bounds}
In this work, we investigate the following conjectural lower bounds on PoQ and qv-OSP.

\begin{conjecture}
    If PoQ exist, then $\BPP \neq \NP$.
\end{conjecture}

\begin{conjecture}
    If qv-OSP exist, then $\BQP \neq \QCMA$.
\end{conjecture}

In particular, we will give oracle evidence that neither of these conjectures hold. This opens up the door to constructing PoQ and qv-OSP from much weaker assumptions than previously expected.

Our oracle separations are significantly stronger than simply resolving the above conjectures, and thus imply a series of independently interesting related separations.

\subsection{Our Results and Corollaries}

\paragraph{PoQ can exist even if $\BPP = \NP$.}

We construct a classical oracle relative to which proofs of quantumness exist, but $\BPP=\NP$. Relative to our oracle, it will also be the case that $\BQP = \PSPACE$. In particular, we show

\begin{theorem}\label{thm:intro-main1}
    There exists a classical oracle relative to which
    \begin{enumerate}
        \item $\P = \NP$
        \item $\BQP = \PSPACE$
        \item There is a problem in $\IP \subseteq \BQP$ which is hard-on-average for $\P/\mathsf{poly}$. 
        \item a polynomial-round public-coin (publicly-verifiable) PoQ exists
        \item a two-message qv-PoQ exists
    \end{enumerate}
\end{theorem} 

Note that nearly all forms of classical cryptography (in particular, everything implying one-way functions) cannot exist if $\P=\NP$. Similarly, nearly all forms of quantum cryptography (in particular, everything implying quantum commitments) cannot exist if $\BQP = \PSPACE$.\footnote{This follows from a $\PSPACE$-effective variant of OpenAI's unitary synthesis algorithm~\cite{OAI26}.} 

\begin{corollary}
    There is no relativizing construction of (most forms of) classical or quantum cryptographic primitives from polynomial-round PoQ or two-message qv-PoQ.
\end{corollary}

Notably, there is no relativizing construction of OSP or PoQM from PoQ, even in the setting of inefficient verification.

As $\P=\NP$, constant-round proofs of quantumness similarly do not exist. This gives a barrier towards round reduction for PoQ as well as a partial barrier for constructing classically verifiable PoQ from qv-PoQ. 

\begin{corollary}
    For every fixed constant $r$, there is no relativizing construction of an $r$-round PoQ from polynomial-round PoQ or two-message qv-PoQ.
\end{corollary}

We remark that for fully black-box constructions, a negative answer for round-reduction is implicit in prior work. By combining two papers~\cite{ITCS:MorYam23,C:ShmZha25} one can construct polynomial-round PoQ from indistinguishability obfuscation and one-way functions. On the other hand, Tomer and Zhandry~\cite{TZ26} rule out fully black-box constructions of constant-round PoQ from these same primitives. 

We also get a partial negative answer to the important open question of whether PoQ implies classical verification of quantum computation (CVQC)~\cite{FOCS:Mahadev18a}. In particular, if a constant-round CVQC protocol existed relative to our oracle, then applying the CVQC to the hard problem for $\P/poly$ gives a constant-round PoQ, which does not exist relative to our oracle.

\begin{corollary}
    For every fixed constant $r$, there is no relativizing construction of an $r$-round CVQC from polynomial-round PoQ or two-message qv-PoQ.
\end{corollary}

We remark that the way we prove that PoQ exists relative to our oracle is by demonstrating a hard problem in $\BQP^\O$ which has an interactive proof whose prover is implementable in $\BQP^\O$. Extracting this fact from our proof gives the following independently interesting corollary.

\begin{corollary}
    Relative to the oracle $\O$ from~\Cref{thm:intro-main1}, there exists a language $L^\O$ such that
    \begin{enumerate}
        \item There is a classically verifiable, quantum computable interactive proof for $L^\O$.
        \item $L^\O$ is hard on average for $\P^\O/poly$
    \end{enumerate}
\end{corollary}

In particular, since $\PH^\O\subseteq \P^\O/poly$ this gives an oracle relative to which there is a problem outside of $\PH$ which can be decided with classical verification of quantum computation.

\paragraph{qv-OSP can exist even if $\BQP = \QCMA$}

We construct a quantum oracle relative to which qv-OSP exists, but $\BQP = \QCMA$.

\begin{theorem}
    There exists a quantum oracle relative to which
    \begin{enumerate}
        \item $\BQP = \QCMA$
        \item a four message qv-OSP exists
    \end{enumerate}
\end{theorem}

The question of $\BQP=\QCMA$ is also closely tied to a quantum cryptographic primitive known as efficiently-verifiable one-way puzzles (EV-OWPuzz)~\cite{C:ChuGolGra24}. The existence of EV-OWPuzzs implies $\BQP \neq \QCMA$. Furthermore, EV-OWPuzzs are implied by a wide variety of quantum cryptographic primitives, including all of post-quantum cryptography. 

Goldin, Morimae, Mutreja, and Yamakawa~\cite{EPRINT:GMMY24} previously showed
that there is no relativizing construction of EV-OWPuzzs from either QCCC commitments or QCCC key exchange. Since qv-OSP implies both of these primitives, Corollary~\ref{cor:ev-owpuzz} strengthens these separations by showing that
even qv-OSP does not suffice for a relativizing construction of EV-OWPuzzs.

Note that many quantum cryptographic primitives are known to imply EV-OWPuzzs~\cite{C:ChuGolGra24}. Selecting interesting examples from this set, we get the following corollary

\begin{corollary}\label{cor:ev-owpuzz}
    For all $A$ in
    \begin{enumerate}
        \item EV-OWPuzzs
        \item post-quantum one-way functions
        \item classically implementable oblivious transfer
        \item quantum public key encryption with classical keys
        \item classically verifiable OSP
    \end{enumerate}
    for all $B$ in
    \begin{enumerate}
        \item four message qv-OSP
        \item qv-PoQM
        \item search-secure QCCC oblivious transfer
    \end{enumerate}
    there is no relativizing construction of $A$ from $B$.
\end{corollary}

Note that the fact that classically verifiable OSP implies EV-OWPuzzs is new. For details, see~\Cref{sec:commitment-ev-puzzles}.

\subsection{Related Works}
\cite{STOC:MorShiYam25} showed that PoQ imply 
either $\mathsf{NP}\not\subseteq\mathsf{BQP}$ or $\mathsf{SampBQP}\neq\mathsf{SampBPP}$. 
A natural question is whether the first alternative alone is necessary: does PoQ imply 
$\mathsf{NP}\not\subseteq\mathsf{BQP}$? Our first result provides an oracle counterexample to this stronger implication, 
because relative to our oracle, PoQ exist but $\mathsf{NP}=\mathsf{P}\subseteq\mathsf{BQP}$.
Thus, the disjunctive necessary condition of \cite{STOC:MorShiYam25} cannot be strengthened to 
$\mathsf{NP}\not\subseteq\mathsf{BQP}$ by a relativizing argument. 

\cite{TZ26} showed that constant-round PoQ imply either two-message PoQ, quantumly-secure OWFs, or (weak) signature token schemes. 
They also show that if polynomial-round PoQ can be constructed from a quantumly-secure falsifiable assumption in a black-box way then quantumly-secure OWPuzz exist.
Although these results are the first derivations of quantum cryptographic primitives from PoQ,
the former is disjunctive and the latter is the derivation from the way of constructing PoQ, not PoQ itself.

The concurrent work \cite{Mantri} showed that relative to a classical oracle there is a language $L$
such that $L\in\mathsf{EQP}\setminus\mathsf{PH}$ and $L$ has an interactive proof system with 
a PPT verifier and an honest QPT prover.
Our results yield an analogous phenomenon for $\mathsf{BQP}$, while additionally providing a PoQ secure against nonuniform classical provers in a world where $\mathsf{P}=\mathsf{NP}$ and $\mathsf{BQP}=\mathsf{PSPACE}$.

\cite{BHNST26} showed a classical oracle relative to which $\mathsf{BQP}\not\subseteq\mathsf{IP}$, 
establishing a relativization barrier to information-theoretically sound, single-prover classical verification of quantum computation. Our result is complementary: we construct a classical oracle relative to which $\mathsf{P}=\mathsf{NP}$ and $\mathsf{BQP}=\mathsf{PSPACE}$, yet polynomial-round proofs of quantumness with efficient classical verification exist. 

\section{Technical Overview}
\subsection{A classical oracle relative to which $\mathsf{P} = \mathsf{NP}$ but proofs of quantumness exist}

We begin with an overview of \Cref{thm:intro-main1}. Prior work on quantum advantage relative to oracles~\cite{Simon,BBBV97,STOC:CCDFGS03,STOC:Aaronson10,STOC:AarAmb15,STOC:RazTal19,AIK22,FOCS:YamZha22,BHNST26} typically proceeds in two steps:

\begin{itemize}
    \item Identify an oracle $\mathcal A$ relative to which a non-trivial quantum algorithm solves an apparently challenging computational problem.
    \item Prove (or conjecture) that classical algorithms relative to that oracle -- potentially augmented by a more powerful oracle such as $\mathsf{PH}^{\mathcal A}$ -- cannot solve the task.
\end{itemize}
One notable extension of this methodology is an approach of Aaronson, Ingram, and Kretschmer~\allowbreak\cite{AIK22}, who use a so-called \emph{Forrelation encoding} to embed the answers to powerful (in their case, $\mathsf{P}^{\#\mathsf{P}}$) computations into an oracle in a form that quantum algorithms can efficiently decode, while the encoding looks random to algorithms in the polynomial hierarchy.

\paragraph{Our methodology: prover-efficient interactive proofs.} Our proof relies on this idea from~\allowbreak\cite{AIK22} but diverges from the above high-level template. We construct a proof of quantumness by giving an efficient prover interactive proof for a decision problem that is hard for $\BPP$ relative to our oracle. That is, the honest prover in our interactive proof can be implemented with a QPT algorithm, and soundness of the interactive proof is information-theoretic.

Let us see under what precise conditions this suffices to get a proof of quantumness. As a warm-up, first suppose that we only want a protocol with soundness error of $1/2 + \epsilon$ against \emph{uniform} $i.o.\mathsf{BPP}$ adversaries. Then, it suffices to work with a \emph{unary} language $L\subseteq\{1\}^*$ outside $\BPP^\O$. To see this, suppose that we have such a language where for each bit $b$, the claim $b=\mathbf{1}_L(1^n)$ has an interactive proof with a QPT honest prover. Then, one gets a proof of quantumness in which the honest quantum prover (on input $1^n$) send the bit $b$ declaring whether $1^n \in L$ and then gives an interactive proof for this fact. The protocol is complete by construction, and satisfies soundness (against classical randomized adversaries) because, by the information theoretical soundness of the interactive proof, any adversary passing the PoQ is deciding membership in $L$. 

To achieve a full-fledged protocol with (standard) soundness against $\mathsf{P/poly}$, we instead work with a (binary) language $L$ that is \emph{non-uniformly hard on average} for a uniformly sampled input $x$. In this case, the proof of quantumness is as follows:

\begin{itemize}
    \item The verifier samples a uniformly random input $x$.
    \item The prover decides whether $x\in L$, sending the corresponding bit $b$ to the verifier.
    \item The prover and verifier execute the interactive proof for $x \in L_b$ (either $L$ or $\overline L$).
\end{itemize}
Again, soundness follows because a prover passing the PoQ is correctly deciding whether a random $x\in L$ by the soundness of the interactive proof. 

\paragraph{Constructing an appropriate oracle with a hard unary language.}
Next, we describe a full solution to the warm-up problem. As mentioned earlier, we deviate from the template of most quantum advantage oracle separations. Why? Because relative to most of the oracles considered in prior work, it is not clear that the problems separating $\mathsf{BQP}$ from classical computation admit efficient-prover interactive proofs. It is actually not clear that they admit interactive proofs of any kind!\footnote{For the complement of Simon's problem, an interactive proof based on approximate counting is known, but Aaronson~\cite[Section~7]{Aaronson21Open} proposed it as a candidate for lacking an interactive proof with a QPT honest prover.} Indeed, relative to a random oracle it is known that even simple classes like $\mathsf{coNP}$ do not have interactive proofs~\cite{FortnowSipser88,CCGHHRR94}, while a recent work~\cite{BHNST26} proves the same for a variant of the Forrelation problem. 

In other words, natural choices of oracle such as a random oracle or an oracle for the Forrelation problem do not seem ``$\mathsf{IP}$-friendly.'' On the other hand, we know that in the \emph{unrelativized} setting, $\mathsf{IP} = \mathsf{PSPACE}$, so this problem arises because this result (famously) does not relativize. 

Our solution is to use Algebrization \cite{STOC:AarWig08} and build our oracle by the following process

\begin{itemize}
    \item First, construct an oracle $\mathcal A$ relative to which $\mathsf{IP} = \mathsf{PSPACE}\supsetneq \mathsf{PH}$, so that there are problems outside of the polynomial hierarchy with interactive proofs.
    \item Then, introduce an oracle $\mathcal B$ so that $\mathsf{PSPACE}^{\mathcal A} \subset \mathsf{BQP}^{\mathcal B}$. This means that every problem in $\mathsf{PSPACE}^{\mathcal A}$ has an interactive proof \emph{with a $\mathsf{BQP}^{\mathcal B}$}-efficient prover!
    \item Finally, prove that there are still problems in $\mathsf{PSPACE}^{\mathcal A}$ that are hard relative to $\mathsf{PH}^{\mathcal A, \mathcal B}$. 
\end{itemize}

Then, by appending an appropriately defined ``$\mathsf{PH}^{\mathcal{A}, \mathcal B}$ oracle,'' we obtain an oracle $\mathcal O = (\mathcal A, \mathcal B, \mathcal C)$ relative to which $\mathsf{P} = \mathsf{NP}$ but a $\mathsf{BQP}$ prover-efficient interactive proof for a classically hard language exists.\footnote{This can be strengthened to an oracle relative to which $\mathsf{BQP} = \mathsf{PSPACE}$ by recursively defining $\mathcal B$ and $\mathcal C$ so that $\mathsf{BQP}$ can solve all of $\mathsf{PSPACE}^{\mathcal O}$.} 

This is an outline of our approach. To complete the warm-up, the final question is how to instantiate $\mathcal A$ and $\mathcal B$. First, we take $\mathcal A = (f, \oplus \mathsf{P}^f)$, where $f$ is a random function. It follows that 
\[\mathsf{PSPACE}^{\mathcal A} = \mathsf{PSPACE}^f \subseteq \mathsf{IP}^{\mathcal A}\subseteq \mathsf{PSPACE}^{\mathcal A}
\]
using the Goldwasser-Kalai-Rothblum variant of the $\mathsf{IP}=\mathsf{PSPACE}$ theorem~\cite{FOCS:Shamir90,STOC:GolKalRot08,STOC:AarWig08,AydinliogluBach17}, which builds an IP for $\mathsf{PSPACE}$ using only low-degree extensions of Boolean functions over characteristic $2$ fields. This is extremely convenient because we know (see, e.g., \cite[Section~3]{Fortnow09}) that
\[ \mathsf{PH}^{\mathcal A} = \mathsf{PH}^{\oplus \mathsf{P}^f} \subseteq \mathsf{BPP}^{\oplus P^f}.
\]
This gives a hard unary language in $\mathsf{PSPACE}^{\mathcal A}$ that is hard for $\mathsf{PH}^{\mathcal A}$ using the majority of the function $f$ over all of its $n$-length inputs. This language is hard for $\mathsf{PH}^{\mathcal A}$ by the $\mathsf{AC}^0[\oplus]$ lower bound for majority \cite{razborov_lower_1987}. 

However, at this point we have an $\mathsf{IP}$ for a language outside $\mathsf{PH}$, but it does not have an \emph{efficient} quantum prover. To do this, we employ an \cite{AIK22} Forrelation encoding of $\mathsf{PSPACE}^{\mathcal A}$, and call this oracle $\mathcal B$. Then, we have that $\mathsf{PSPACE}^{\mathcal A}$ is contained in $\mathsf{IP}^{\mathcal A}$ with a $\mathsf{BQP}^{\mathcal B}$-efficient prover. On the other hand, by the security of the Forrelation encoding, we have that $\mathsf{PH}^{\mathcal A, \mathcal B}$ algorithms cannot solve majority on $f$ with probability much better than a $\mathsf{PH}$ algorithm relative to $\mathcal A$ along with an independent random oracle. This completes the warm-up. 

\paragraph{Extending to non-uniform hardness.}
Of course, the unary language used in our warm-up PoQ is not hard for non-uniform classical adversaries, which can have memorized the language in their advice strings. To get around this, we naturally turn to \emph{salting}: for a function $f: \{0,1\}^n \times \{0,1\}^n \rightarrow \{0,1\}$, we define the language
\[ L^f = \{x \in \{0,1\}^n : \sum_{y\in\{0,1\}^n} f(x,y)>2^{n-1} \}.
\]
For a random function $f$, this gives $2^n$ independent majority instances at each input length, indexed by a uniformly sampled $x$. Then, by constructing the same oracle $\mathcal O = (\mathcal A, \mathcal B, \mathcal C)$ we get our candidate oracle relative to which $\mathsf{P} = \mathsf{NP}$ and there is an efficient-prover interactive proof for the language $L^f$. 

The remaining question is how to show that deciding $L^f$ is almost everywhere hard-on-average for $(\mathsf{PH}/\mathsf{poly})^{\mathcal A, \mathcal B}$. Fix a universal oracle Turing machine $M$ and an advice length $s(n)=\poly(n)$. We want to upper bound the quantity
\[
    \rho^\delta(\mathcal A, \mathcal B) = \Pr_{\mathcal A, \mathcal B}\!\left[\exists z\in\{0,1\}^{s(n)}:\quad
        \Pr_{x\gets\{0,1\}^n}[M^{\mathcal O}(x,z)=\mathrm{maj}(f_x)]\geq 1-\delta\right].
\]
Our argument proceeds as follows. Suppose instead that this quantity is large. 

\begin{enumerate}
    \item \emph{Carefully invoke security of the Forrelation encoding.} As a first step (which turns out to be the hardest part of the proof), we would like to convert this attack into one where the Forrelated $\mathsf{PSPACE}$ oracle $\mathcal B$ has been replaced by a random oracle, invoking the classical ($\mathsf{PH}$) hardness of Forrelation. Compared to before, the challenge is handling advice strings $z$ that can depend on the oracles $\mathcal A, \mathcal B$; since these can reveal individual bits of the Forrelation encoding, it's not immediately clear how to invoke the hardness of Forrelation.

    In other words, we would like to argue that the quantity $\rho(\mathcal A, \mathcal B)$ does not change by much when we replace $\mathcal B$ by a random oracle. If $\rho(\mathcal A, \mathcal B)$ were computable by a $\mathsf{PH}^{\mathcal A, \mathcal B}$ machine, then we would be done. Unfortunately, this doesn't seem to be the case:
    \begin{itemize}
        \item The verification itself involves the quantity $\mathrm{maj}(f_x)$, which is hard to compute, and
        \item The expectation over $x$ is also not naturally a $\mathsf{PH}$ computation.
    \end{itemize}
    To fix this, we introduce a \emph{proxy}
    \[ \rho_t^\delta(\mathcal A, \mathcal B, x_1, \hdots, x_t, y_1, \hdots, y_t) = \Pr_{\mathcal A, \mathcal B}\!\left[\exists z\in\{0,1\}^{s(n)}:\quad
        \Big| i: M^{\mathcal O}(x_i,z)=y_i \Big|\geq 1-\delta\right].
    \]
    By standard concentration arguments, if $t \gg s$ then we have that
    \[   \underset{x_1, \hdots, x_t}{\mathbb E} \rho_t^{\delta/2}(\mathcal A, \mathcal B, x_1, \hdots, x_t, \mathrm{maj}(f_{x_1}), \hdots, \mathrm{maj}(f_{x_t})) < \rho^\delta(\mathcal A, \mathcal B) + \epsilon
    \]
    and
    \[\rho^\delta(\mathcal A, \mathcal B) <  \underset{x_1, \hdots, x_t}{\mathbb E} \rho_t^{2\delta}(\mathcal A, \mathcal B, x_1, \hdots, x_t, \mathrm{maj}(f_{x_1}), \hdots, \mathrm{maj}(f_{x_t})) +\epsilon
    \]
    for a small error $\epsilon$. That is, if we sample random $x_1, \hdots, x_t$ and precompute the corresponding majority values $y_i = \mathrm{maj}(f_{x_i})$, then the quantity $\rho(\mathcal A, \mathcal B)$ that we care about can be approximated by the proxy with small losses in the parameters. Crucially, the proxy is efficiently computable by a $\mathsf{PH}^{\mathcal A, \mathcal B}$ machine.

    Finally, we invoke Forrelation hardness to argue that the quantity 
    \[\underset{x_1, \hdots, x_t}{\mathbb E} \rho_t(\mathcal A, \mathcal B, x_1, \hdots, x_t, \mathrm{maj}(f_{x_1}), \hdots, \mathrm{maj}(f_{x_t}))
    \]
    is approximately preserved when switching $\mathcal B$ to a random oracle.\footnote{This is arguing that the Forrelation encoding can be used to hide a ``message'' sampled from a fixed distribution depending on $\mathcal A$ but not the randomness of the Forrelation oracle.} Combining these arguments, we can conclude that if $\rho^\delta(\mathcal A, \mathcal B)$ is noticeably large in the case where $\mathcal B$ is Forrelated, then $\rho^{O(\delta)}(\mathcal A, \mathcal B)$ is noticeably large when $\mathcal B$ is a random oracle. 

    \item \emph{Reuse the advice to solve many instances.} Now that $\mathcal B$ has been replaced by a random oracle, relatively standard techniques for handling adversaries with advice can be applied. Specifically, suppose that a non-uniform $\mathsf{PH}^{\mathcal A, \mathcal B}$ adversary computes $\mathrm{maj}(f_x)$ with probability $1-\delta$. Then, if we ran the adversary -- reusing the advice string $z$ -- on independent inputs $x_1, \hdots, x_t$, we would immediately obtain that the adversary succeeds on all $t$ inputs with probability $(1-\delta)^t$. In turn, this implies a \emph{uniform} $\mathsf{PH}$ algorithm for the $t$-fold problem that succeeds with probability $2^{-s} \cdot (1-\delta)^t$ by guessing the advice string $z$ uniformly.

    \item \emph{Uniform hardness using a direct-product theorem.} In the random oracle-circuit complexity correspondance, the above algorithm gives a quasi-polynomial size $\mathsf{AC}^0[\oplus]$ circuit that computes $t$ simultaneous majorities on independent random inputs with probability $2^{-s} \cdot (1-\delta)^t$. The fact that such circuits do not exist follows from Razborov's majority lower bound combined with an $\mathsf{AC}^0$-friendly direct-product theorem~\cite{razborov_lower_1987,product}. 
\end{enumerate}

These steps, combined with standard union bound arguments, yields a proof that salted majority is weakly hard-on-average, which gives us a proof of quantumness with constant soundness error. This soundness error can then be amplified to negligible by standard repetition. 
\subsection{An oracle relative to which $\BQP=\QCMA$ but oblivious state preparation exists}

For our second result, we construct a quantum (unitary) oracle relative to which $\BQP=\QCMA$, but the (very strong) quantum primitive of quantum verifier oblivious state preparation (qv-OSP) exists. Similarly to~\cite{Kretschmer21,EPRINT:GMMY24}, we will construct an oracle $\O$ relative to which it is easy to construct qv-OSP secure against inefficient, query-bounded adversaries, but for which efficiently identifiable information about the oracle cannot be encoded classically. In particular, it will be the case that for any function $c(\O)$ output a polynomial length classical string, it will be impossible for an oracle algorithm $\A^{(\cdot),\PSPACE}(c(\O))$ to tell if it is interacting with the oracle $\O$ or a freshly sampled oracle $\wt{\O}$\footnote{
Note that this will only hold for the portions of the oracle acting on sufficiently large inputs. Oracles with small inputs will be handled using the same techniques as~\cite{Kretschmer21}.}. i.e. with high probability over the choice of $\O$,
$$\abs{\Pr[\A^{\O,\PSPACE}(c(\O))\to 1] - \Pr_{\wt{\O}}[\A^{\wt{\O},\PSPACE}(c(\O)) \to 1]} = \negl$$

As such, for $\PSPACE$ an oracle to a $\PSPACE$ complete problem, relative to $\O,\PSPACE$ qv-OSP exists. This is because the $\PSPACE$ oracle is independent of $\O$, and so cannot help inefficient, query-bounded algorithms. However, any $\QCMA^{\O,\PSPACE}$ verifer $\Ver^{\O,\PSPACE}$ will have approximately the same behavior on all inputs when $\O$ is replaced with a fresh random oracle $\wt{\O}$ sampled by the verifier itself. As such, for all $\Ver^{\O,\PSPACE}$ there is an equivalent verifier $\wt{\Ver}^{\PSPACE}$ deciding the same language and making no queries to $\O$. Therefore, the language corresponding to $\Ver^{\O,\PSPACE}$ can be decided in $\BQP^{\PSPACE}$ by checking if the modified $\wt{\Ver}^{\PSPACE}$ accepts.

For simplicity, we will first construct a quantum channel oracle, and then describe how we lift this to a unitary oracle.

\paragraph{Our quantum channel oracle}

Our oracle will be indexed by collections of Haar random states $\ket{\phi_{x}}$ for $x\in \{0,1\}^n$, as well as a random oracle $\RO_n:\{0,1\}^n \times \{0,1\}^{4n} \to \{0,1\}^{6n}\times \{0,1\}^{6n}$. For an input $x,y$, write $\RO_n(x,y) = u_{x,y},v_{x,y}$.
\begin{enumerate}
    \item $\SG_n$: Samples a random $x\in \{0,1\}^n$. Outputs $(x,\ket{\phi_x})$.
    \item $\Expand_n(x)$: Samples a random $y\in \{0,1\}^{4n}$. Outputs $y$ as well as the "claw state"
    $$\ket{c_{x,y}} \propto \ket{u_{x,y}} + \ket{v_{x,y}}$$
    \item $\Mix_n(x,y,\ket{\psi})$: If $\ket{\psi} = \ket{\phi_x}$, outputs $\ket{\RO_n(x,y)} = \ket{u_{x,y}}\ket{v_{x,y}}$.
\end{enumerate}

\paragraph{Constructing qv-OSP}

To construct qv-OSP, it suffices to construct a "claw-state generator"~\cite[Theorem~4.7]{C:BarKhu25}. This is a classical communication interactive protocol where at the end of the protocol the verifier receives two strings $a,b$ while the prover receives the "claw-state" $\frac{1}{\sqrt{2}}\ket{a} + \frac{1}{\sqrt{2}}\ket{b}$. Security says that at the end of the protocol, a malicious prover cannot predict both $a$ and $b$ with better than negligible probability.

Constructing a claw-state generator relative to our oracle is straightforward. The protocol is as follows
\begin{enumerate}
    \item The verifier will run $\SG_n\to (x,\ket{\phi_x})$ and sends $x$ to the prover
    \item The prover will run $\Expand_n(x) \to (y,\ket{c_{x,y}})$. It will send $y$ to the verifier, and output the claw state $$\ket{c_{x,y}} \propto \ket{u_{x,y}} + \ket{v_{x,y}}.$$
    \item Finally, the verifier will query $\Mix_n(x,y,\ket{\phi_x}) \to (u_{x,y},v_{x,y})$ and output $a = u_{x,y},b=v_{x,y}$.
\end{enumerate}

Correctness of the protocol is clear. Security comes from the fact that the prover does not see $\ket{\phi_x}$, and so cannot access $\Mix(x,\cdot)$. As such, its only access to $\RO_n(x,\cdot)$ is through $\Expand_n(x)$. In particular, it can only see one copy of $\ket{c_{x,y}}$ for each $y$ randomly sampled by $\Expand_n(x)$. Thus, the only way it can predict both $u_{x,y},v_{x,y}$ simultaneously for any $y$ is if it guesses $\ket{\phi_x}$,  guesses $u_{x,y},v_{x,y}$ directly, or $\Expand_n$ samples the same $y$ twice. These events occur with at most negligible probability.

In addition, we prove that security holds against non-uniform adversaries with classical advice on the oracle. This will follow by essentially repeating the same techniques used to show $\BQP^{\O,\PSPACE} = \QCMA^{\O,\PSPACE}$, which essentially says that classical information on the oracle is useless.

\paragraph{Proving $\BQP^{\O,\PSPACE} = \QCMA^{\O,\PSPACE}$}

As described above, this proof boils down to showing with high probability over the choice of $\O$,
\begin{equation}\label{eq:tech-overview-useless}
\abs{\Pr[\A^{\O,\PSPACE}(c(\O))\to 1] - \Pr_{\wt{\O}}[\A^{\wt{\O},\PSPACE}(c(\O)) \to 1]} = \negl
\end{equation}
That is, it is impossible to describe any efficiently identifiable information about $\O$ using polynomially many classical bits.

Note that there are two components of $\O$ one could hope to describe: the quantum states $\ket{\phi_x}$ and the random oracle $\RO$. It is well known that a short amount of classical information cannot encode any identifiable information about a Haar random state~\cite{Kretschmer21}. This is known as superconcentration of the Haar distribution.

And so, it remains to show that a short classical string cannot give any identifiable information about the random oracle. Just as in~\cite{EPRINT:GMMY24}, the key here is that any algorithm $\A$ can only access the random oracle at \textit{random points}. In particular, at points $\RO_n(x,\cdot)$ for $x$ sampled by $\SG_n$ or at points $\RO_n(\cdot,y)$ for $y$ sampled by $\Expand_n$. However, by the presampling technique~\cite{EC:CDGS18}, any classical string describing the random oracle can only describe a polynomial number of its inputs. Thus, any classical advice $c(\O)$ is unlikely to carry any information about $\RO_n(x,\cdot),\RO_n(\cdot,y)$ for sampled $x,y$. The claim in~\Cref{eq:tech-overview-useless} follows.

\paragraph{Lifting to a unitary oracle}

In general, while unitary oracle separations can be difficult to reason about, there are a number of "stronger" oracle types which are easier to work with. One particularly useful oracle formulation are state preparation channels. These are oracles defined by some family of mixed states $\sigma_x$. When queried on an input $x$, the oracle will return a copy of $\sigma_x$. Such oracles have recently been used to achieve a number of separations in quantum cryptography~\cite{TCC:AnaGulLin24,EC:CheColSat25}.

One can also consider state preparation \textit{unitaries}. A state preparation unitary for a mixed state $\sigma$ is any unitary $S_{\sigma}$ such that
$$\sigma_{\Reg{A}} = \Tr_{\Reg{B}}(S_{\sigma}\ketbra{0}_{\Reg{A}\Reg{B}} S_{\sigma}^\dagger)$$

A series of recent works has given techniques for simulating state preparation unitaries with state preparation channels~\cite{EC:BMMMY25,C:GolZha25,EPRINT:GMMY24,EPRINT:AnaGol25,TWZ26}. The culmination of these results gives that for any mixed state $\sigma$, there exists a distribution over state preparation unitaries $S_{\sigma}$ that can be efficiently simulated using copies of $\sigma$ (i.e. by querying a state preparation channel). More formally, there exists simulator $\Sim(\sigma^{\otimes L})$ taking in a polynomial number $L$ of copies of $\sigma$ such that for all polynomial-query oracle algorithms $\A$ outputting a single bit,
$$\abs{\Pr_{S_\sigma}[\A^{S_\sigma}\to 1] - \Pr[\A^{\Sim(\sigma^{\otimes L})}\to 1]} \leq \frac{1}{\poly}$$

An immediate consequence of this is that if a complexity collapse holds relative to a random $S_{\sigma}$, then it also holds relative to a state preparation channel for $\sigma$. It is tempting to then claim that anytime we achieve an oracle separation relative to a state preparation oracle, we can translate it to a unitary oracle separation in this way. However, this is false. In particular, the issue lies in the other direction. If a complexity collapse holds relative to a state preparation oracle for $\sigma$, it need not necessarily also hold relative to $S_{\sigma}^U$.

The key observation implicit in~\cite{EPRINT:GMMY24} is that this reverse direction does hold specifically for the case of $\BQP^{\O,\PSPACE} = \QCMA^{\O,\PSPACE}$. The argument takes advantage of the fact that breaking $\QCMA$ is a non-interactive game using only classical communication.

For a family $\vec{\sigma}=\{\sigma_t\}_{t\in\N}$ of $t$-qubit mixed states, write $\BQP^{\vec{\sigma}},\QCMA^{\vec{\sigma}}$ for the corresponding classes with access to a state preparation channel for each $\sigma_t$. Write $S=\{S_{\sigma_t}^{U_t}\}_{t\in\N}$, where the $U_t$ are independent Haar unitaries on $t$ qubits. It turns out that by combining the ability to simulate $S$ with the techniques of~\cite{Kretschmer21}, one can show $\QCMA^{S} \subseteq \BQP^{S,\QCMA^{\vec{\sigma}}}$. As an immediate corollary, if $\BQP^{\vec{\sigma}} = \QCMA^{\vec{\sigma}}$, then with probability $1$, $\BQP^{S} = \QCMA^{S}$. For full formality, we include a proof of this in \Cref{sec:stateprepcollapse}.

As such, these techniques show that any oracle where $\BQP=\QCMA$ and some cryptographic object exists\footnote{Technically, this argument only applies to "one-stage" cryptographic games, i.e. games where there is only one stateful adversary.} relative to a state preparation channel immediately lifts to a unitary oracle. Note that our oracle is not a state preparation channel oracle. However, our oracle morally makes controlled queries to a channel oracle, and combining with the fact that the individual prepared states are themselves simulatable, similar techniques apply.

\section{Preliminaries}\label{sec:prelims}

\subsection{Notation}

For $m\in\N$, write $[m]=\{1,\dots,m\}$. For a finite set $X$, $x\gets X$ denotes a uniformly random element of $X$; for a distribution $\D$, $x\gets\D$ denotes a sample from $\D$. All logarithms are base $2$ unless written as $\ln$. For a finite set $X$, let $\mathcal H(X)$ be the Hilbert space with orthonormal basis $\{\ket{x}:x\in X\}$. Write $\mathrm U(D)$ for the group of unitaries on a $D$-dimensional space. We use $\Haar(X)$ for the Haar distribution on unitaries on $\mathcal H(X)$, and $\HaarSt(X)$ for the Haar distribution on pure states in $\mathcal H(X)$.

We write $\poly(n)=n^{O(1)}$, $\polylog(n)=(\log n)^{O(1)}$, and $\quasipoly(n)=2^{(\log n)^{O(1)}}$ for unspecified bounds of these forms. A nonnegative function $\nu(n)$ is negligible if for every constant $c>0$, $\nu(n)\leq n^{-c}$ for all sufficiently large $n$; we write $\negl(n)$ for an unspecified negligible function.

In acceptance-probability bounds, algorithms output a single bit, and the probability includes all internal randomness and measurements. For a discrete random variable $X$, its min-entropy is $H_\infty(X)=-\log\max_x\Pr[X=x]$.

\subsection{Complexity Preliminaries}\label{sec:complexity-prelims}

\subsubsection{Algorithms and circuits}
PPT and QPT denote probabilistic and quantum polynomial-time algorithms, respectively. Quantum algorithms use a fixed finite universal gate set, computational basis measurements, and ancillas initialized to $\ket{0}$. Uniform circuit families have descriptions generated by a deterministic polynomial-time algorithm. Nonuniform algorithms are polynomial-size circuit families; their descriptions, or equivalently their classical advice, may depend on the input length and the oracle. We do not allow quantum advice unless stated explicitly. Query-bounded algorithms need not be time bounded unless this is stated.

We measure the size of a circuit by its description length in bits. Under our convention, there are at most $2^{\ell}$ circuits of size $\ell$.

\subsubsection{Oracle Algorithms}\label{sec:oracle-algorithms}
A classical oracle is a function $f$ on bit strings. A classical oracle algorithm may submit an input $x$ and receive $f(x)$, choosing each query using the answers to previous queries. Preparing and storing queries and reading answers count towards the algorithm's time and space bounds.

A quantum oracle is a family of unitary operators. We allow controlled queries and queries to both a unitary and its inverse. Components and levels may be queried in superposition, using zero padding for unused registers. A query to level $t$ takes at least $t$ time. A classical function $f$ is accessed quantumly through the map $\ket{x,z}\mapsto\ket{x,z\oplus f(x)}$. All random choices in our oracle constructions are independent unless a dependence is specified.

\subsubsection{Oracle complexity classes}
We identify a language $L\subseteq\{0,1\}^*$ with its characteristic function $L:\{0,1\}^*\to\{0,1\}$. A superscript denotes oracle access. In the classical definitions below, $\O$ is a classical oracle. In the quantum definitions, $\U$ is a quantum oracle with the query conventions of~\Cref{sec:oracle-algorithms}, following~\cite{Kretschmer21}. Classical oracles are included as a special case, and multiple oracles may be listed in a superscript.

\begin{definition}[$\P^\O$]\label{def:oracle-p}
    A language $L$ is in $\P^\O$ if there is a deterministic polynomial-time oracle Turing machine $M^\O$ such that, for every $x\in\{0,1\}^*$,
    \[
        M^\O(x)=L(x).
    \]
\end{definition}

\begin{definition}[$\NP^\O$]\label{def:oracle-np}
    A language $L$ is in $\NP^\O$ if there are a deterministic polynomial-time oracle verifier $\Ver^\O$ and a polynomial $p$ such that, for every $x\in\{0,1\}^*$,
    \[
        x\in L\quad\Longleftrightarrow\quad
        \exists w\in\{0,1\}^{p(|x|)},\quad \Ver^\O(x,w)=1.
    \]
\end{definition}

\begin{definition}[$\oplus\P^\O$]\label{def:oracle-parity-p}
    A language $L$ is in $\oplus\P^\O$ if there are a deterministic polynomial-time oracle verifier $\Ver^\O$ and a polynomial $p$ such that, for every $x\in\{0,1\}^*$,
    \[
        L(x)=\bigoplus_{w\in\{0,1\}^{p(|x|)}}\Ver^\O(x,w).
    \]
    Thus, $x\in L$ exactly when the number of accepting witnesses is odd.
\end{definition}

\begin{definition}[$\BPP^\O$]\label{def:oracle-bpp}
    A language $L$ is in $\BPP^\O$ if there is a PPT oracle algorithm $\A^\O$ such that, for every $x\in\{0,1\}^*$,
    \[
        \Pr[\A^\O(x)\to L(x)]\geq 2/3.
    \]
\end{definition}

\begin{definition}[$\PSPACE^\O$]\label{def:oracle-pspace}
    A language $L$ is in $\PSPACE^\O$ if there is a deterministic oracle Turing machine $M^\O$ which halts on every input, uses polynomial space including its query tape, and satisfies $M^\O(x)=L(x)$ for every $x\in\{0,1\}^*$.
\end{definition}

\begin{definition}[$\BQP^\U$]\label{def:oracle-bqp}
    A language $L$ is in $\BQP^\U$ if there is a QPT oracle algorithm $\A^\U$ such that, for every $x\in\{0,1\}^*$,
    \begin{enumerate}
        \item If $x\in L$, then $\Pr[\A^\U(x)\to 1]\geq 2/3$.
        \item If $x\notin L$, then $\Pr[\A^\U(x)\to 1]\leq 1/3$.
    \end{enumerate}
\end{definition}

\begin{definition}[$\QCMA^\U$]\label{def:oracle-qcma}
    A language $L$ is in $\QCMA^\U$ if there are a QPT oracle verifier $\Ver^\U$ and a polynomial $p$ such that, for every $x\in\{0,1\}^*$,
    \begin{enumerate}
        \item \textbf{Completeness.} If $x\in L$, there is a classical witness $w\in\{0,1\}^{p(|x|)}$ such that\\
        $\Pr[\Ver^\U(x,w)\to 1]\geq 2/3$.
        \item \textbf{Soundness.} If $x\notin L$, then for every $w\in\{0,1\}^{p(|x|)}$, $\Pr[\Ver^\U(x,w)\to 1]\leq 1/3$.
    \end{enumerate}
\end{definition}

\begin{definition}[$\IP^\O$]\label{def:oracle-ip}
    A language $L$ is in $\IP^\O$ if there is a PPT oracle verifier $V^\O$ in a polynomial-round protocol with classical communication such that, for every $x\in\{0,1\}^*$,
    \begin{enumerate}
        \item \textbf{Completeness.} If $x\in L$, there is an unbounded prover $P^\O$ such that\\
        $\Pr[(V^\O\leftrightarrows P^\O)(x)\to 1]\geq 2/3$.
        \item \textbf{Soundness.} If $x\notin L$, then for every unbounded prover $P^{*\O}$, $\Pr[(V^\O\leftrightarrows P^{*\O})(x)\to 1]\leq 1/3$.
    \end{enumerate}
    Both parties have access to $\O$.
\end{definition}

Omitting the oracle gives the corresponding unrelativized classes. A complexity class in an oracle superscript denotes a fixed complete problem for that class. For a promise problem $(L_{\mathrm{yes}},L_{\mathrm{no}})$, the same acceptance and verification conditions are required only on promised inputs: replace $x\in L$ by $x\in L_{\mathrm{yes}}$ and $x\notin L$ by $x\in L_{\mathrm{no}}$. We specify explicitly when complexity classes refer to promise problems.

\subsubsection{Circuit classes}
\begin{definition}[Circuit classes]\label{def:circuit-classes}
We write $\AC^0(S,d)$ for the class of Boolean circuits of size at most $S$ and depth at most $d$, using unbounded fan-in AND and OR gates and fan-in one NOT gates. The class $\AC^0[\xor](S,d)$ also allows unbounded fan-in parity gates. The classes $\AC^0$ and $\AC^0[\xor]$ consist of nonuniform families of polynomial-size, constant-depth circuits of these forms. Size follows the convention above, and depth is the largest number of gates on an input-to-output path. An oracle gate evaluates the indicated oracle on the bits supplied to that gate.
\end{definition}

\subsubsection{Multilinear Extensions}\label{sec:multilinear-extensions}
We first define an important concept known as the multilinear extension of a boolean function to a finite field. In particular, let $f:\{0,1\}^t\to\{0,1\}$ be any function. We define the multilinear extension of $f$ over $\F_{2^d}^t$ by
\begin{equation}
    \ML_{2^d}[f](x) \coloneqq \sum_{u\in \{0,1\}^t} f(u)\chi_{u}(x)\quad \chi_u(x)\coloneqq \prod_{i=1}^t (u_ix_i + (1 - u_i)(1-x_i))
\end{equation}
Note that $\ML_{2^d}[f]$ is a multilinear function (i.e. it is of degree at most $1$ in each variable and thus of total degree at most $t$). Furthermore, viewing $\{0,1\}^t$ as a subset of $\F_{2^d}^t$ under the natural inclusion, we have for all $x\in \{0,1\}^t$ that $\chi_u(x)=1$ if $x=u$ and $\chi_u(x) = 0$ if $x\neq u$. Thus, for all $x\in \{0,1\}^t$, $\ML_{2^d}[f](x) = f(x)$.

\subsubsection{Circuit Lower Bounds}\label{sec:circuit-lower-bounds}

\paragraph*{Forrelation}
We give the following statement, which says that there exists a problem (Forrelation) which can be efficiently solved quantumly but which cannot be solved by quasipolynomial size $\AC^0$ circuits.

\begin{lemma}[Hardness of Forrelation, Theorem 1.2 of~\cite{STOC:RazTal19}]\label{lem:forr}
    There exists an explicit distribution $\mathcal{F}_{N}$ (Forrelation) over $\{0,1\}^N$ such that 
    \begin{enumerate}
        \item there exists a quantum algorithm $\A$ making $\polylog(N)$ queries to its input and running in time $\polylog(N)$ such that:
        $$\abs{\Pr_{x\gets \mathcal{F}_N}[\A(x) \to 1] - \Pr_{x\gets \{0,1\}^N}[\A(x)\to 1]} \geq 1-\frac{1}{N^2}$$
        \item For any $C\in \AC^0(\quasipoly(N),O(1))$:
        $$\abs{\Pr_{x\gets \mathcal{F}_N}[C(x) \to 1] - \Pr_{x\gets \{0,1\}^N}[C(x)\to 1]} \leq \frac{\polylog(N)}{\sqrt{N}}$$
    \end{enumerate}
\end{lemma}

Note that the hardness of Forrelation also applies even when $C$ is additionally given as input samples from any distribution independent of $x$.

\paragraph*{Hardness of Majority}
Let $\MAJ:\{0,1\}^*\to\{0,1\}$ be the function such that $\MAJ(x)$ is $1$ if and only if more than half the bits of $x$ are $1$.

We state the following lower bound on the ability for $\AC^0$ circuits to compute the majority function. This lower bound comes originally from~\cite{razborov_lower_1987}. For an explanation, see~\cite{kopparty_lecture}. The precise statement we use appears as Theorem 3 of~\cite{Hoza2026}.

\begin{theorem}[Majority lower bound]\label{thm:majoritylb}
    Let $\MAJ(x)$ be the majority function. Then for any $C\in \AC^0[\xor](S,O(1))$
    $$\Pr_{x\gets \{0,1\}^N}[C(x) = \MAJ(x)] \leq \frac{1}{2} + \frac{\polylog(S)}{\sqrt{N}}$$
\end{theorem}

\paragraph*{Direct product theorem for circuits}

\begin{theorem}[Direct product theorem for circuits (Theorem 1.2 of~\cite{product})]\label{thm:directproduct}
    There exists a universal constant $c>0$ such that the following holds. Let $\delta\in (0,1)$, $f:\{0,1\}^n\to \{0,1\}$ be any boolean function, and $\Samp$ any distribution over $\{0,1\}^n$. Let $\A$ be any algorithm such that
    $$\Pr_{\Samp\to x_1,\dots,x_t}[\A(x_1,\dots,x_t) = (f(x_1),\dots,f(x_t))] \geq e^{-c\delta t}$$
    Then there exists an $\AC^0$ circuit $C$ with oracle gates of size $\poly(n,t,\log(1/\delta),e^{c\delta t})$ such that
    $$\Pr_{\Samp\to x}[C^\A(x) = f(x)] \geq 1-\delta$$
\end{theorem}
\subsection{Techniques for Simulating Oracles}\label{sec:oracle-simulation-techniques}

\subsubsection{Unitary Tomography}

\begin{theorem}[Unitary tomography~\cite{FOCS:HKOT23}]\label{thm:tomography}
    Let $U$ be a unitary on $m$ qubits and let $\epsilon,\delta\in(0,1)$. There is a quantum algorithm with controlled query access to $U$ which runs in time $\poly(2^m,\allowbreak1/\epsilon,\allowbreak\log(1/\delta))$ and outputs a classical description of a unitary circuit $\wt U$. With probability at least $1-\delta$, for every $q\in\N$ and every oracle algorithm $\A^{(\cdot)}$ making at most $q$ queries and outputting a single bit,
    \[
        \abs{\Pr[\A^U\to 1]-\Pr[\A^{\wt U}\to 1]}\leq q\epsilon.
    \]
    Queries may be controlled, forwards or inverse, and the bound allows arbitrary auxiliary registers and fixed side oracles.
\end{theorem}

\begin{remark}
    Unitary tomography as stated by~\cite{FOCS:HKOT23} specifically says that one can learn a description of a unitary $\wt{U}$ which is close to $U$ under a metric known as the diamond norm. A basic induction then gives that $\wt{U}$ is indistinguishable from $U$ by bounded query algorithms, where the error blows up linearly in the number of queries.

    A second subtlety emerges in that the guarantee only applies to algorithms which do not make controlled queries to their oracle. However, applying tomography instead to the controlled version of $U$ 
    $$\ketbra{0}\otimes I+\ketbra{1}\otimes U$$
    gives the above theorem.
\end{remark}

\subsubsection{Reprogramming Unitaries}

We will use as well a simplified statement of the unitary reprogramming lemma from~\cite{CGR26,TCC:HhaYam25}. This says that unitary oracles which differ on unpredictable points are indistinguishable.

\begin{lemma}[Unitary reprogramming lemma]\label{lem:reprogramming}
    Let $\epsilon \in (0,1)$. Let $\Pi\gets S$ be a distribution over projectors such that for any fixed state $\ket{\phi}$,
    $$\E_{\Pi\gets S}\left[\norm{\Pi \ket{\phi}}^2\right] \leq \epsilon$$

    Let $\O$ be any unitary oracles, and let $\ol{\O}_\Pi$ be any family of oracles such that
    $$\O (I - \Pi) = \ol{\O}_\Pi(I-\Pi)$$

    Then for all $q$ query algorithms $\D^{(\cdot)}$, we have
    $$\abs{\Pr[\D^{\O} \to 1] - \Pr_{\Pi\gets S}[\D^{\ol{\O}_{\Pi}} \to 1]}\leq 2q\sqrt{\epsilon}$$
\end{lemma}

\begin{theorem}[Random unitaries are useless~\cite{Kretschmer21,Mec19}]\label{thm:unitarysim}
    Let $D\in \N$, $\epsilon\in (0,1)$ and let $\A^{(\cdot)}$ be any algorithm making $q$ queries to its oracle. Then
    $$\Pr_{U\gets \Haar([D])}\left[\abs{\Pr[\A^U \to 1] - \Pr_{U' \gets \Haar([D])}[\A^{U'}\to 1]} \geq \epsilon \right] \leq 2\exp\left(-\frac{(D-2)\epsilon^2}{96q^2}\right)$$
\end{theorem}

\subsubsection{Simulating State Preparation}

\begin{lemma}[Replacing unitaries with states~\cite{EPRINT:GMMY24,TCC:AnaGulLin24}]
    Let $N<M\in \N$. Let $\rho$ be the mixed state of dimension $M^N$ consisting of $N$ random elements of $[M]$. That is,
    $$\rho = \E_{k_i\gets [M],i\in [N]}\left[\bigotimes_{i=1}^N \ketbra{k_i}\right]$$

    Let $\rho'$ be the same but we replace $\ket{k_i}$ with $\ket{\phi_i}$, a Haar random state subject to the restriction that for each $i\neq j$, $\ket{\phi_i},\ket{\phi_j}$ are orthogonal. That is,
    $$\rho' = \E_{U\gets \Haar([M])}\left[\bigotimes_{i=1}^N U\ketbra{i}U^\dagger\right]$$

    Then, for every quantum algorithm $\A$ outputting a single bit,
    $$\abs{\Pr[\A(\rho)\to 1]-\Pr[\A(\rho')\to 1]} \leq \frac{N(N+1)}{2M}.$$
\end{lemma}

\begin{remark}
    Although Lemma 2.24 of~\cite{EPRINT:GMMY24} uses independent Haar random states, since a single copy of a Haar random state is maximally mixed it is sufficient to consider random strings instead.
\end{remark}

\begin{definition}[State preparation for pure states]
    Let $\ket{\phi}$ be a pure state. The state preparation unitary for $\ket{\phi}$ is the map which swaps $\ket{0}\ket{0}$ and $\ket{1}\ket{\phi}$ defined by
    $$S_{\ket{\phi}} = I - \ketbra{1}\otimes \ketbra{\phi} - \ketbra{0}\otimes \ketbra{0} + \ketbra{0}{1} \otimes \ketbra{0}{\phi} + \ketbra{1}{0} \otimes \ketbra{\phi}{0}$$
\end{definition}

\begin{definition}[State preparation for mixed states]
    Let $\sigma$ be a mixed state. We say that a map $S_{\sigma}$ is a state preparation unitary for $\sigma$ if there exists a pure state $\ket{\phi}_{AB}$ such that $\sigma = \Tr_B(\ketbra{\phi}_{AB})$ and $S_{\sigma} = S_{\ket{\phi}}$.
\end{definition}

\begin{definition}[Random state preparation]
    Let $\sigma$ be a mixed state on $\mathcal{H}(\{0,1\}^n)$. Fix any purification $\ket{\phi}\in \mathcal{H}(\{0,1\}^n)\otimes \mathcal{H}(\{0,1\}^m)$ with $\sigma = \Tr_B(\ketbra{\phi}_{AB})$. Let $S_{\sigma} = S_{\ket{\phi}}$ be the state preparation unitary for $\sigma$. Let $U \in \mathrm U(2^m)$. We define $S_{\sigma}^U$ to be the map
    \begin{align*}
        S_{\sigma}^U &= S_{(I_{\mathcal{H}(\{0,1\}^n)} \otimes U)\ket{\phi}} = C_U S_{\sigma} C_U^\dagger,\\
        C_U &= \ketbra{0}\otimes I + \ketbra{1}\otimes (I_{\mathcal{H}(\{0,1\}^n)}\otimes U).
    \end{align*}
\end{definition}

\begin{lemma}[Simulating state preparation unitaries~\cite{TWZ26}]\label{lem:stateprep}
    There exists a universal constant $c>0$ such that the following holds. Let $q,n,m\in \N$, let $\epsilon\in (0,1)$, and let $L\geq cq^2/\epsilon$. Then there exists a stateful algorithm $\Sim$ with the following property. Let $\sigma$ be any mixed state on $\mathcal{H}(\{0,1\}^n)$ with purification $\ket{\phi} \in \Hilb{n} \otimes \Hilb{m}$ and $S_\sigma = S_{\ket{\phi}}$. Then for all oracle algorithms $\A^{(\cdot)}$ making at most $q$ (possibly controlled, forwards or inverse) queries and outputting a single bit, we have
    $$\abs{\Pr_{U\gets \Haar(\{0,1\}^m)}[\A^{S_{\sigma}^U}\to 1] - \Pr[\A^{\Sim(\sigma^{\otimes L})}\to 1]} \leq \epsilon$$
    
    Furthermore, there is an efficient circuit implementing $\Sim$ of size $\poly(L,n,m)$.
\end{lemma}

\begin{lemma}[Simulating families of state preparation unitaries]\label{cor:stateprep}
    There exists a universal constant $c>0$ such that the following holds. Let $q,\ell,n,m\in \N$, let $\epsilon\in (0,1)$, and let $L\geq cq^2/\epsilon$. Then there exists a stateful algorithm $\Sim$ with the following property. Let $\{\sigma_x\}_{x\in [\ell]}$ be a family of mixed states on $\mathcal{H}(\{0,1\}^n)$ with purifications $\ket{\phi_x} \in \Hilb{n} \otimes \Hilb{m}$ and $S_{\sigma_x} = S_{\ket{\phi_x}}$. Define $\Sigma = \bigotimes_{x\in [\ell]} \sigma_x^{\otimes L}$.
    For a family $\{U_x\}_{x\in [\ell]}$, define $S^{\{U_x\}_{x\in [\ell]}} = \sum_x \ketbra{x} \otimes S_{\sigma_x}^{U_x}$. Then for all oracle algorithms $\A^{(\cdot)}$ making at most $q$ (possibly controlled, forwards or inverse) queries and outputting a single bit, we have
    $$\abs{\Pr_{U_x\gets \Haar(\{0,1\}^m),\,x\in[\ell]}[\A^{S^{\{U_x\}}}\to 1] - \Pr[\A^{\Sim(\Sigma)}\to 1]} \leq \epsilon$$
    
    Furthermore, there is an efficient circuit implementing $\Sim$ of size $\poly(L, \ell,n,m)$.
\end{lemma}

\begin{proof}
    As in the proof of~\Cref{lem:stateprep}, apply the procedures from~\cite[Lemmas~2.3 and~2.11]{TWZ26} to the copies of each $\sigma_x$. This gives $L$ copies of a random pure state $\ket{\eta_x}$ such that
    $$S_{\sigma_x}^{U_x}=I-2P_x,\qquad P_x=\ketbra{\eta_x}.$$
    Here the $U_x$ are independent Haar unitaries, with the same $U_x$ used for all copies of $\ket{\eta_x}$. The random phases introduced by the procedures can be absorbed into the $U_x$. Define
    $$H=\sum_x\ketbra{x}\otimes P_x,$$
    so the oracle we want to simulate is $I-2H=e^{-i\pi H}$.

    Let $B$ be the target register of a query. Take one unused copy of each $P_x$ in registers $E_x$, and define
    $$\omega_E=\bigotimes_xP_x,\qquad V=\sum_x\ketbra{x}\otimes\mathrm{SWAP}_{B,E_x}.$$
    Thus, $V$ swaps $B$ with the copy indexed by $x$, without measuring $x$. Applying $e^{-ihV}$ and tracing out $E$ gives the channel
    $$\Phi_h(\rho)=\Tr_E\left[e^{-ihV}(\rho\otimes\omega_E)e^{ihV}\right]
    =\rho-ih[H,\rho]+O(h^2).$$
    Since $\norm{V}_{\infty}=1$ and $\norm{H}_{\infty}\leq 1$, the error term changes acceptance probabilities by at most $O(h^2)$, even in the presence of other registers. The map $\rho\mapsto e^{-ihH}\rho e^{ihH}$ has the same first-order expansion, so one such operation simulates it up to error $O(h^2)$. This bound also holds when the label register is in superposition or entangled with other registers.

    For each query, repeat this operation $r=\floor{L/q}$ times using unused copies and $h=\pi/r$. The error for one query is $O(rh^2)=O(1/r)$, and so the total error over $q$ queries is $O(q/r)=O(q^2/L)\leq\epsilon$ for sufficiently large $c$. This uses at most $qr\leq L$ copies for each $x$. For controlled queries, replace $H$ and $V$ by $\ketbra{1}\otimes H$ and $\ketbra{1}\otimes V$. Inverse queries are the same as forward queries since the oracle is a reflection. Finally, averaging over the $U_x$ preserves the error bound, and the procedures above can be implemented to the required accuracy using $\poly(L,\ell,n,m)$ gates.
\end{proof}

\subsubsection{Presampling Lemmas}

\begin{theorem}[Presampling~\cite{EC:CDGS18}]
    Let $\mathcal{Y}$ be any finite set. We define a partial assignment $P = \{(x_1,y_1),\allowbreak\dots,\allowbreak(x_L,y_L)\}$ to be a subset of $\{0,1\}^n \times \mathcal{Y}$ such that the $x_i$ are distinct. Write $\Dom(P)=\{x:\exists y,\,(x,y)\in P\}$. For a function $f$, define a new function $f[P]$ by $f[P](x) = y$ for all $(x,y)\in P$, otherwise $f[P](x) = f(x)$.

    Let $n,q,L,s\in \N$ with $1\leq L\leq 2^n$, and let $\epsilon\in (0,1)$. Let $\D^{(\cdot)}$ be an oracle algorithm making at most $q$ classical queries to its oracle. Let $c:\{f:\{0,1\}^n\to \mathcal{Y}\}\to \{0,1\}^s$ be any function, and let $f,\wt{f}$ be independent uniformly random functions. Then there exists a family $\{P_z\}_{z\in \{0,1\}^s}$ of distributions over partial assignments of length $L$ such that
    $$\abs{\Pr_{f}[\D^f(c(f)) \to 1] - \Pr_{\substack{f,\wt{f}\\P\gets P_{c(f)}}}[\D^{\wt{f}[P]}(c(f)) \to 1]} \leq \frac{(s+\log(1/\epsilon))\cdot q}{L}+\epsilon$$
\end{theorem}

\begin{corollary}[Advice can't distinguish exponentially many random query points]\label{cor:indis}
    Let $\mathcal{Y}$ be any finite set, and let $f,\wt{f}:\{0,1\}^n\to\mathcal{Y}$ be independent uniformly random functions. Let $X_1,\dots,X_q$ be random variables over $\{0,1\}^{n}$, jointly independent of $f,\wt{f}$, such that for each $i$, $H_{\infty}(X_i) \geq \gamma$. Let $c:\{f:\{0,1\}^n\to \mathcal{Y}\}\to \{0,1\}^s$ be any function. Let $\A$ be any (potentially inefficient or quantum) randomized algorithm. Then
    $$\abs{\Pr_{f}[\A(c(f),f(X_1),\dots,f(X_q))\to 1] - \Pr_{f,\wt{f}}[\A(c(f),\wt{f}(X_1),\dots,\wt{f}(X_q))\to 1]} \leq 2^{-n} + \frac{(s+n+1)q}{2^{\gamma/2}}$$
    The same bound holds if $\A$ also receives the sampled points $X_1,\dots,X_q$.
\end{corollary}

\begin{proof}
    The claim is trivial if the stated bound is at least $1$, so assume otherwise; in particular, $s+n\geq 2$. Consider the distinguisher $\D^f(c(f))$ which samples $X_1,\dots,X_q$, queries $f(X_1),\dots,f(X_q)$, and then outputs $\A(c(f),f(X_1),\dots,f(X_q))$. This is an algorithm making at most $q$ classical queries to its oracle. By setting $\epsilon = 2^{-n}$ and $L = \ceil{2^{\gamma/2}}$, we have
    $$\begin{aligned}
        &\abs{\Pr_{f}[\A(c(f),f(X_1),\dots,f(X_q)) \to 1] - \Pr_{\substack{f,\wt{f}\\P\gets P_{c(f)}}}[\A(c(f),\wt{f}[P](X_1),\dots,\wt{f}[P](X_q)) \to 1]}\\
        &\qquad\leq \frac{(s+n)q}{L} + 2^{-n}
    \end{aligned}$$

    But note that if none of $X_1,\dots,X_q$ lie in $\Dom(P)$, then $\wt{f}[P](X_i) = \wt{f}(X_i)$ is a fresh uniformly random value. And by the definition of min-entropy and the union bound, since $P$ is of size $L$, we have that the probability that any $X_i$ lies in $\Dom(P)$ is $\leq \frac{qL}{2^\gamma}$. Since $2^{\gamma/2}\leq L\leq 2\cdot 2^{\gamma/2}$ and $s+n\geq 2$, we have
    $$\frac{s+n}{L}+\frac{L}{2^\gamma}\leq\frac{s+n+1}{2^{\gamma/2}}.$$
    And so
    $$\abs{\Pr_{f}[\A(c(f),f(X_1),\dots,f(X_q)) \to 1] - \Pr_{\substack{f,\wt{f}}}[\A(c(f),\wt{f}(X_1),\dots,\wt{f}(X_q)) \to 1]} \leq 2^{-n} + \frac{(s+n+1)q}{2^{\gamma/2}}$$
    Giving the sampled points to $\A$ does not change the number of queries made by $\D$ or the probability of hitting $P$, so the same proof applies.
\end{proof}

\subsection{Cryptographic Preliminaries}\label{sec:crypto-prelims}

\subsubsection{Proofs of Quantumness}

\begin{definition}[Proof of quantumness~\cite{FOCS:BCMVV18,STOC:AGKZ20}]\label{def:poq}
    A proof of quantumness relative to $\O$ is a protocol with classical communication between a QPT prover $P^\O$ and a PPT verifier $V^\O$, on common input $1^n$. The verifier outputs a bit, and the following properties hold:
    \begin{enumerate}
        \item \textbf{Completeness.} The honest prover makes the verifier accept with probability
        $$\Pr[(V^\O\leftrightarrows P^\O)(1^n)\to 1]\geq 1-\negl(n).$$
        \item \textbf{Classical soundness.} For every nonuniform PPT classical prover $P^{*\O}$,
        $$\Pr[(V^\O\leftrightarrows P^{*\O})(1^n)\to 1]\leq\negl(n).$$
    \end{enumerate}
    The probabilities are over the parties' randomness and measurements. Allowing $V^\O$ to be QPT instead gives a quantumly-verifiable proof of quantumness.
\end{definition}

\begin{remark}
    A proof of quantumness with a classical verifier is \emph{public coin} if every verifier message consists of fresh uniformly random bits, and the verifier's final decision is a deterministic function of $1^n$, the transcript, and $\O$~\cite[Definition~7.1]{STOC:AGKZ20}.
\end{remark}

\subsubsection{Oblivious State Preparation}

\begin{definition}[Oblivious state preparation~\cite{C:BarKhu25}]\label{def:osp}
    Oblivious state preparation relative to $\O$ is a protocol with classical communication between a PPT verifier $V^\O$ and a QPT prover $P^\O$, on common input $1^n$. The verifier also has an input bit $b$, outputs a bit $s$, and the prover outputs one qubit. It satisfies:
    \begin{enumerate}
        \item \textbf{Correctness.} For each $b\in\{0,1\}$, measuring the honest prover's output in the computational basis if $b=0$, or the Hadamard basis if $b=1$, gives $s$ with probability at least $1-\negl(n)$.
        \item \textbf{Security.} For every nonuniform QPT malicious prover $P^{*\O}$ outputting a bit,
        \[
            \abs{\Pr[P^{*\O}\to 1\text{ in }\langle V^\O(1^n,0),P^{*\O}(1^n)\rangle]
            -\Pr[P^{*\O}\to 1\text{ in }\langle V^\O(1^n,1),P^{*\O}(1^n)\rangle]}\leq\negl(n).
        \]
    \end{enumerate}
    Allowing the verifier to be QPT instead gives quantum verifier oblivious state preparation (qv-OSP).
\end{definition}

\cite{C:BarKhu25} defines a notion of claw state generators, which are sufficient to construct oblivious state preparation. Intuitively, a claw state generator lets a (classical) verifier produce two strings $a,b$ while giving a quantum prover the claw $\frac{1}{\sqrt{2}}\left(\ket{a}+\ket{b}\right)$. Security says that the quantum prover should not be able to produce \textit{both} $a$ and $b$. We thus define an oracle version of claw state generators, and show how to construct them relative to our oracle. Note that unlike~\cite{C:BarKhu25}, we allow the verifier to be quantum, although we restrict communication to be classical.

\begin{definition}[Claw state generator]\label{def:claw-state-generator}
    A \emph{claw state generator} relative to $\O$ is a protocol with classical communication between quantum polynomial-time parties $\mathsf{V}^{\O}$ and $\mathsf{P}^{\O}$ on input $1^n$.
    The verifier outputs $(a,b)\in\{0,1\}^{6n}\times\{0,1\}^{6n}$ or aborts, and the honest prover outputs a $6n$-qubit register.
    The protocol satisfies:
    \begin{enumerate}
        \item \textbf{Correctness.} Let $\rho$ be the joint output state of an honest execution, and define
        \[
            \ket{\chi_{a,b}} = \frac{\ket{a}+\ket{b}}{\sqrt{2}}
            \qquad (a\ne b),
        \]
        \[
            \Pi_{\mathrm{CSG}} =
            \sum_{\substack{a,b\in\{0,1\}^{6n},\ a\ne b}}
            \ket{a,b}\bra{a,b}
            \otimes\ket{\chi_{a,b}}\bra{\chi_{a,b}}.
        \]
        Then $\Tr(\Pi_{\mathrm{CSG}}\rho)\geq 1-\negl(n)$.
        \item \textbf{Search security.} For every nonuniform quantum polynomial-time malicious prover $\mathsf{P}^{*\O}$,
        \[
            \Pr\!\left[
                \{a',b'\}=\{a,b\}
                \;:\;
                \bigl((a,b),(a',b')\bigr)
                \gets\langle\mathsf{V}^{\O}(1^n),\mathsf{P}^{*\O}(1^n)\rangle
            \right]\leq\negl(n).
        \]
        The event is false if the verifier aborts; the probability includes all protocol randomness and measurements.
    \end{enumerate}
\end{definition}

\begin{theorem}[Adapted from Theorem~4.7 of~\cite{C:BarKhu25}]\label{thm:csg-implies-osp}
    For all oracles $\O$, if a claw-state generator exists relative to $\O$, then qv-OSP exists relative to $\O$.
\end{theorem}

\begin{remark}
    Theorem 4.7 of~\cite{C:BarKhu25} states that if claw-state generators with classical verification imply oblivious state preparation with classical verification. We remark that their proof immediately extends to quantum verifiers when we allow the oblivious state preparation protocol to use quantum verification. Their argument is black-box and thus relativizes.
\end{remark}

\section{A classical oracle with quantum advantage but no cryptography}

It is known that there exists an oracle where $\P^\O = \NP^\O$ and $\BQP^\O = P^{\PP^\O}$~\cite{AIK22}. The oracle consists of random oracle, a (recursively defined) $\NP$ oracle, as well as a (recursively defined) quantum accessible $P^{\PP}$ oracle. In particular, classical queries to the latter oracle are prevented by encoding the oracle under the Forrelation problem. For outputs where the oracle would output $1$, instead a pair of Forrelated random oracles are provided, and for outputs where the oracle would output $0$, instead a pair of independent random oracles are provided.

By replacing $P^{\PP}$ with $\PSPACE$, the same proof gives an oracle where $\P^\O=\NP^\O$, $\BQP^\O=\PSPACE^\O$, and $\NP^\O \subsetneq \BQP^\O$. If these complexity collapses held in the standard model (i.e. without oracles), we would get the following template for a proof of quantumness:

Pick a hard problem in $\BQP$. Since $\BQP \subseteq \IP$, there is an interactive proof for this problem. Since $\BQP = \PSPACE$, the interactive proof can be implemented by a QPT algorithm. And so the proof of quantumness will consist of the verifier choosing a problem instance, sending it to the prover, who then answers the problem and gives an interactive proof that its answer is correct. This can be implemented quantumly, but since the problem is hard and the proof is sound it cannot be implemented classically.

There are two main problems with this template. The first issue is that $\PSPACE \subseteq \IP$ (or even $\BQP \subseteq \IP$) does not relativize~\cite{BHNST26}. In particular, it may be the case that $\IP^\O \subsetneq \PSPACE^\O$ relative to the oracle defined by~\cite{AIK22}. The second problem is that the hard problem given by~\cite{AIK22} is only infinitely often worst-case hard against uniform adversaries. If we want a proof of quantumness to be cryptographically secure, we need a much stronger notion: average case hardness against non-uniform adversaries for all sufficiently large security parameters. In this section, we give an oracle resolving both of these issues.

\begin{theorem}
    There exists a distribution over triples of classical oracles $\O = (\OracleA,\OracleB,\OracleC) \gets \D$ such that with probability $1$,
    \begin{enumerate}
        \item $\P^\O = \NP^\O$
        \item $\BQP^\O = \PSPACE^\O$
        \item $\IP^{\OracleA} = \PSPACE^{\OracleA}$
        \item There exists a constant $\delta \in (0,1)$ and a boolean function $f:\{0,1\}^*\to \{0,1\}$ such that
        \begin{enumerate}
            \item for all non-uniform PPT algorithms $\A$, for all sufficiently large $n$,
                $$\Pr_{x\gets \{0,1\}^n}[\A^\O(x) = f(x)] \leq 1-\delta$$
            \item the language $L^f = \{x\in \{0,1\}^* : f(x) = 1\}$ is in $\PSPACE^{\OracleA} \subseteq \IP^\O$
        \end{enumerate} 
    \end{enumerate}
\end{theorem}

\subsection{Our Oracle}

Similarly to the proof of Theorem 32 in~\cite{AIK22}, we define a distribution $\D$ over classical oracles $\O = (\OracleA,\OracleB,\OracleC) \gets \D$ consisting of three parts. We will have $\OracleA = \{\OracleA_t\}_{t\in \N}, \OracleB = \{\OracleB_t\}_{t\in \N}, \OracleC=\{\OracleC_t\}_{t\in \N}$, where $\OracleB_t$ and $\OracleC_t$ will be defined inductively. We will write $\O_t = (\OracleA_t,\OracleB_t,\OracleC_t)$.

Let $\RO_t:\{0,1\}^{t} \to \{0,1\}$ be a random oracle. We will write $\RO = \{\RO_t\}$.  Define $\OracleA_t(d,x) = \ML_{2^d}[\RO_t](x)$ to be the multilinear extension of $\RO$.

Now, we will describe $\OracleB_t$ and $\OracleC_t$, which will enforce that $\BQP^\O = \PSPACE^\O$ and $\P^\O = \NP^\O$ respectively.

Denote $S_t^{\PSPACE}$ to be the set of ordered pairs $(M,x)$ such that
\begin{enumerate}
    \item $M$ is an oracle Turing machine and $x$ is an input to $M$.
    \item $(M,x)$ takes less than $t$ bits to specify
    \item $M$ is syntactically restricted in the following three ways
    \begin{enumerate}
        \item $M$ runs in at most $2^t$ steps
        \item $M$ uses space at most $t$
        \item $M$ queries only the oracles $\{\O_1,\dots,\O_{\floor{\sqrt{t}}}\}$
    \end{enumerate}
\end{enumerate}

For each $(M,x) \in S_t^{\PSPACE}$, sample the function $\OracleB_t((M,x),\cdot):\{0,1\}^{t^2}\to \{0,1\}$ as follows
\begin{enumerate}
    \item If $M^\O(x)$ accepts, then sample $\OracleB_t((M,x),\cdot)$ from the Forrelation distribution $\mathcal{F}_{2^{t^2}}$
    \item If $M^\O(x)$ rejects, then sample $\OracleB_t((M,x),\cdot)$ uniformly at random
\end{enumerate}
Thus, $\OracleB_t:S_t^{\PSPACE} \times \{0,1\}^{t^2}\to \{0,1\}$.

Define $S_t^{\NP}$ to be the set of ordered pairs $(M,x)$ such that
\begin{enumerate}
    \item $M$ is a nondeterministic Turing machine and $x$ is an input to $M$.
    \item $(M,x)$ takes less than $t$ bits to specify
    \item $M$ is syntactically restricted to run in at most $t$ steps and query only the oracles $(\O_1,\allowbreak\dots,\allowbreak\O_{\floor{\sqrt{t}}})$
\end{enumerate}
Define $\OracleC_t:S_t^{\NP} \to \{0,1\}$ to be
$$\OracleC_t(M,x) = \begin{cases}
    1 & M^\O(x)\text{ accepts}\\
    0 & \text{otherwise}
\end{cases}$$
In order to enforce that a query to $\OracleB_t,\OracleC_t$ requires algorithms to run in time $t$, we will have $\OracleB_t,\OracleC_t$ ignore padding as well as inputs of shorter than $t$ bits. A query to $\OracleB_t,\OracleC_t$ on an input shorter than $t$ bits we will implicitly take to mean a query on that input padded to $t$ bits.

\subsection{Complexity classes collapse under our oracle}

We will show the following three important complexity collapses relative to our oracle (or some subset thereof).

\begin{lemma}\label{lem:pnp}
    Let $\O = (\OracleA,\OracleB,\OracleC)$ be any oracle in the support of $\D$, then
    $$\P^\O = \NP^\O$$
\end{lemma}

\begin{proof}
    Given an oracle $\NP$ machine $M$ running in time at most some computable polynomial $p(|x|)$ and querying only the oracles $(\O_1,\dots,\O_{p(|x|)})$, a $\P^\O$ machine can produce $M^\O(x)$ by outputting $\OracleC_t(M,x)$ for $t=\max(|M,x|+1,p(|x|)^2)$ (where $|M,x|$ is the number of bits needed to specify $M$ and $x$).
\end{proof}

\begin{lemma}\label{lem:bqppspace}
    With probability $1$ over $\O = (\OracleA,\OracleB,\OracleC)\gets\D$,
    $$\BQP^\O = \PSPACE^\O$$
\end{lemma}

\begin{proof}
    We follow the proof of Claim 30 in~\cite{AIK22}. Let $Q^{\OracleB_t}(M,x)$ be the quantum algorithm from~\Cref{lem:forr} applied to $\OracleB_t((M,x),\cdot)$, with output $1$ corresponding to the Forrelation distribution. For every $(M,x)\in S_t^{\PSPACE}$, its error averaged over $\O\gets\D$ is at most $2^{-2t^2}$. By Markov's inequality and a union bound over the at most $2^t$ pairs in $S_t^{\PSPACE}$,
    \[
        \Pr_{\O\gets\D}\left[\exists(M,x)\in S_t^{\PSPACE}:\Pr_Q[Q^{\OracleB_t}(M,x)\neq M^\O(x)]>1/3\right]
        \leq 3\cdot 2^{t-2t^2}.
    \]
    Since $\sum_t 3\cdot 2^{t-2t^2}<\infty$, Borel-Cantelli gives that with probability $1$, for all sufficiently large $t$, $Q$ has error at most $1/3$ on every pair in $S_t^{\PSPACE}$.

    Fix such an oracle and a $\PSPACE^\O$ machine $M$. Choose a polynomial $p$ such that $M^\O(x)$ uses at most $p(|x|)$ space, runs in time at most $2^{p(|x|)}$, queries only $\O_1,\dots,\O_{p(|x|)}$, and $|M,x|<p(|x|)$. Set $t=p(|x|)^2$. Then $(M,x)\in S_t^{\PSPACE}$ and $Q^{\OracleB_t}(M,x)$ runs in polynomial time and computes $M^\O(x)$ with error at most $1/3$ for all sufficiently long $x$. Hardcoding the answers on the finitely many remaining inputs gives $\PSPACE^\O\subseteq\BQP^\O$. The reverse inclusion holds for every classical oracle.
\end{proof}

\paragraph*{$\PSPACE^\OracleA = \IP^\OracleA$}

Aaronson and Wigderson~\cite{STOC:AarWig08} show that for any oracle $R$, relative to a low-degree encoding $\wt{R}$ of $R$ over a sufficiently large prime field, $\PSPACE^{R} \subseteq \IP^{\wt{R}}$. The way this argument works is by "algebrizing" the existing argument that $\PSPACE=\IP$ proved by Shamir~\cite{FOCS:Shamir90}. In particular, Shamir's argument gives an interactive proof for the $\PSPACE$ complete language Quantified Boolean Formula ($QBF$) by utilizing the low-degree extension of the corresponding formulae. Since $QBF^R$ is complete for $\PSPACE^R$, one would like to similarly utilize low-degree extensions of formulae querying the oracle $R$. Since the verifier can compute these using access to the low-degree extension of $R$, we get $\PSPACE^R \subseteq \IP^{\wt{R}}$.

While~\cite{STOC:AarWig08} does not immediately apply to our oracle, since it requires the field to have large prime characteristic, Ayd{\i}nl{\i}o{\u g}lu and Bach~\cite{AydinliogluBach17} consider fields of characteristic $2$ directly. They show that $\PSPACE\subseteq\IP$ holds relative to every affine oracle. These are oracles of the same form as $\OracleA$, giving access to the multilinear extensions of a family of Boolean functions over $\F_{2^d}$ for every $d\geq 1$, with field elements represented by bit strings.

\begin{theorem}[Corollary~34 of~\cite{AydinliogluBach17}]\label{thm:affineip}
    Let $\{f_n:\{0,1\}^n\to \{0,1\}\}_{n\in \N}$ be any family of functions. Define $\wt{f} = \{\ML_{2^d}[f_n]\}_{d,n\in \N}$ to be the collection of multilinear extensions of $f_n$ over $\F_{2^d}$, with field elements represented by bit strings. Each extension has total degree at most $n$. Then
    $$\PSPACE^{\wt{f}} \subseteq \IP^{\wt{f}}$$
\end{theorem}

As our oracle $\OracleA$ is an oracle of this form and as $\IP \subseteq \PSPACE$ relativizes, we get the following corollary.

\begin{corollary}\label{lem:ippspace}
    Let $\O = (\OracleA,\OracleB,\OracleC)$ be any oracle in the support of $\D$, then
    $$\IP^\OracleA = \PSPACE^\OracleA$$
\end{corollary}

\begin{remark}
    We emphasize that we explicitly do not show that $\IP^\O=\PSPACE^\O$ relativizes to the entire oracle $\O$; it may be the case that $\IP^\O\subsetneq\PSPACE^\O$.
\end{remark}

\subsection{Representing oracle Turing machines as circuits}

We begin by describing how to represent polynomial time oracle machines by constant depth circuits. We first restate Lemma 35 from~\cite{AIK22}, which shows how to represent $\NP^{\O}$ Turing machines by $\AC^0$ circuits. Although their choice of oracle differs slightly from ours ($\OracleB$ encodes $\PSPACE$ instead of $\P^{\#\P}$ and $\OracleA$ gives multilinear extensions instead of direct access), the lemma only relies on the recursive definition of $\OracleC$ in terms of $\OracleA$ and $\OracleB$, and so can be adapted to our oracle without changing the proof.

\begin{lemma}[Representing oracle Turing machines as $\AC^0$ circuits]\label{lem:ac0}
    Fix $t,d\in \N$ and let $t'\leq t^{2^d}$. For each $(M,x)\in S_{t'}^{\NP}$, there exists an $\AC^0(2^{1+t^{2^d}},2d)$ circuit that takes as input $x$, $(\OracleA_1,\dots,\OracleA_{t^{2^{d-1}}})$, $(\OracleB_1,\dots,\OracleB_{t^{2^{d-1}}})$, $(\OracleC_1,\dots,\OracleC_t)$, and outputs $M^\O(x)$ for all choices of oracle $\O = (\OracleA,\OracleB,\OracleC)$ in the support of $\D$.
\end{lemma}

\begin{proof}
    This follows from the exact same argument as Lemma 35 in~\cite{AIK22}.
\end{proof}

We note that since $\AC^0[\xor]$ circuits can compute $\OracleA$ from $\RO$, we get the following corollary.

\begin{corollary}\label{cor:ac02}
    Fix $t,d\in \N$ and let $t'\leq t^{2^d}$. For each $(M,x)\in S_{t'}^{\NP}$, there exists an $\AC^0[\xor](2^{1+t^{2^d}},\allowbreak 2d+2)$ circuit that takes as input $x$, $(\RO_1,\dots,\RO_{t^{2^{d-1}}})$, $(\OracleB_1,\dots,\OracleB_{t^{2^{d-1}}})$, $(\OracleC_1,\dots,\OracleC_t)$, and outputs $M^\O(x)$ for all choices of oracle $\O$ in the support of $\D$.
\end{corollary}

\begin{proof}
    Note that for each $x\in \F_{2^i}^{j}$ there is a depth $1$ $\AC^0[\xor]$ circuit which computes $\ML_{2^i}[\RO_j](x)$ when given input $\RO_j$. This is because each output bit of $\ML_{2^i}[\RO_j](x)$ is a linear function of the coordinates of $\RO_j$, and so can be computed by taking a parity of some subset of the oracle positions.
    
    Replacing the $\OracleA_i$ input wires of the circuit from~\Cref{lem:ac0} with the results of this computation gives an $\AC^0[\xor]$ circuit with the input scheme given in the lemma. Note that this increases the depth by at most $1$
\end{proof}
\subsection{Forrelation is useless for classical computation}

In this section, we will prove the following main theorem, which states that from the perspective of any classical algorithm the oracle $\OracleB$ can be replaced with truly random functions.

Fix any $n \in \N$. Define $\D^{(n)}$ to be the distribution over oracles defined as follows:
\begin{enumerate}
    \item For $t<n$, sample $\OracleA_t,\OracleB_t,\OracleC_t$ the same way as in $\D$
    \item For $t\geq n$, produce $\OracleA_t,\OracleB_t,\OracleC_t$ exactly the same as when sampling from $\D_t$ with one key difference: always sample $\OracleB_t((M,x),\cdot)$ from the uniform distribution.
\end{enumerate}

Note that the proofs of~\Cref{lem:ac0,cor:ac02} do not depend on the choice of $\RO,\OracleB$, but only on the structure of $\OracleA$ and $\OracleC$. And so they apply just as well to oracles sampled from $\D^{(n)}$.

\begin{lemma}[Forrelated oracle is useless to classical algorithms]\label{lem:classicalforr}
    Let $p(n),q(n),s(n)$ be three polynomials. Let $f_n^\OracleA:\{0,1\}^{s(n)}\to \{0,1\}^{q(n)}$ be any family of functions dependent only on the $\OracleA$ portion of the oracle. Let $\Samp^\OracleA(1^n)\to x\in \{0,1\}^{s(n)}$ be any distribution dependent only on the $\OracleA$ portion of the oracle. Let $M^\O$ be any oracle Turing machine running in time $p(|x|)$. Let $n \in \N$. Then
    $$\abs{\Pr_{\substack{\O=(\OracleA,\OracleB,\OracleC)\gets \D\\x\gets \Samp^\OracleA(1^n)}}[M^{\O}(x)=f_n^\OracleA(x)] - \Pr_{\substack{\O=(\OracleA,\OracleB,\OracleC)\gets \D^{(n)}\\x\gets \Samp^\OracleA(1^n)}}[M^{\O}(x)=f_n^\OracleA(x)]} \leq 2^{-\Omega(n^2)}$$
\end{lemma}

\begin{proof}
    Let $T(n)=p(s(n))$. This is a polynomial bound on the running time of $M^\O$ on inputs of length $s(n)$.

    First, observe that 
    $$\Pr_{\substack{\O=(\OracleA,\OracleB,\OracleC)\gets \D\\x\gets \Samp^\OracleA(1^n)}}[M^{\O}(x)=f_n^\OracleA(x)] = \Pr_{\substack{\O=(\OracleA,\OracleB,\OracleC)\gets \D^{(T(n)+1)}\\x\gets \Samp^\OracleA(1^n)}}[M^{\O}(x)=f_n^\OracleA(x)]$$
    since $M^\O$ does not run in time large enough to query $\O_{T(n)+1}$. We will then proceed to show that for all $i\in [n,T(n)]$, 
    $$\Pr_{\substack{\O=(\OracleA,\OracleB,\OracleC)\gets \D^{(i)}\\x\gets \Samp^\OracleA(1^n)}}[M^{\O}(x)=f_n^\OracleA(x)] \approx \Pr_{\substack{\O=(\OracleA,\OracleB,\OracleC)\gets \D^{(i+1)}\\x\gets \Samp^\OracleA(1^n)}}[M^{\O}(x)=f_n^\OracleA(x)]$$

    For $i\in [n,T(n)]$, we enumerate all pairs in $S_i^{\PSPACE}$ as $(M_{i,1},x_{i,1}),\dots,(M_{i,s_i},x_{i,s_i})$, where $s_i\leq 2^i$. Define $H^{(i)}_j$ to be the same as $\D^{(i+1)}$ but where $\OracleB_i((M_{i,1},x_{i,1}),\cdot),\dots,\OracleB_i((M_{i,j},x_{i,j}),\cdot)$ are sampled uniformly at random instead of from the Forrelation distribution.

    Now, fix any $i,j$. Let $t=i$ and $t'=T(i)^2$. Note that there exists a constant $d$ such that $t' \leq t^{2^d}$. Set $N=2^{i^2}$. By~\Cref{lem:ac0} there is an $\AC^0(\quasipoly(N),O(1))$ circuit $\mathcal{C}$ which takes in as input $x$, $(\OracleA_1,\dots,\OracleA_{t^{2^{d-1}}})$, $(\OracleB_1,\dots,\OracleB_{t^{2^{d-1}}})$, $(\OracleC_1,\dots,\OracleC_t)$ and outputs $M^\O(x)$. Now consider the circuit $\mathcal{C}'$ which additionally takes as input $y=f_n^\OracleA(x)$ and outputs $1$ if and only if $M^\O(x)=y$. This can also be implemented in $\AC^0(\quasipoly(N),O(1))$ by running $\mathcal{C}$ and comparing the output with the input $y$.

    Conditioned on $M_{i,j}^\O(x_{i,j})=0$, $H^{(i)}_j$ and $H^{(i)}_{j-1}$ are identically distributed. And so define $H^*_1,H^*_2$ to be $H^{(i)}_j$ and $H^{(i)}_{j-1}$ conditioned on $M_{i,j}^\O(x_{i,j})=1$. Observe that the only difference between $H^*_1$ and $H^*_2$ is that in $H^*_1$, $\OracleB_i((M_{i,j},x_{i,j}),\cdot)$ is uniformly random and in $H^*_2$, $\OracleB_i((M_{i,j},x_{i,j}),\cdot)$ is sampled from $\mathcal{F}_{N}$.

    Further observe that since $i = t > \sqrt{t}$, when drawn from either of $H^*_1,H^*_2$ all portions of the input to $\mathcal{C}'$ besides $\OracleB_i((M_{i,j},x_{i,j}),\cdot)$ are independent of $\OracleB_i((M_{i,j},x_{i,j}),\cdot)$. And so by~\Cref{lem:forr},
    \begin{equation}
    \begin{split}
        &\Bigg|\Pr_{\O \gets H^{(i)}_j}[M^\O(x) = f_n^\OracleA(x)]-\Pr_{\O \gets H^{(i)}_{j-1}}[M^\O(x) = f_n^\OracleA(x)]\Bigg|\\
        \leq &\Bigg|\Pr_{\O \gets H^*_1}[M^\O(x) = f_n^\OracleA(x)]-\Pr_{\O \gets H^*_2}[M^\O(x) = f_n^\OracleA(x)]\Bigg|\\
        \leq& \Bigg|\Pr_{\O \gets H^*_1}[\mathcal{C}'(x,f_n^\OracleA(x),(\OracleA_1,\dots,\OracleA_{t^{2^{d-1}}}),(\OracleB_1,\dots,\OracleB_{t^{2^{d-1}}}),(\OracleC_1,\dots,\OracleC_t)) \to 1]\\
        &-\Pr_{\O \gets H^*_2}[\mathcal{C}'(x,f_n^\OracleA(x),(\OracleA_1,\dots,\OracleA_{t^{2^{d-1}}}),(\OracleB_1,\dots,\OracleB_{t^{2^{d-1}}}),(\OracleC_1,\dots,\OracleC_t)) \to 1]\Bigg|\\
        \leq& \frac{\polylog(N)}{\sqrt{N}} = \frac{\poly(i)}{2^{i^2/2}}
    \end{split}
    \end{equation}

    Summing over all $i\in[n,T(n)],j\in[1,s_i]$, we get
    \begin{equation}
        \begin{split}
            &\abs{\Pr_{\substack{\O=(\OracleA,\OracleB,\OracleC)\gets \D\\x\gets \Samp^\OracleA(1^n)}}[M^{\O}(x)=f_n^\OracleA(x)] - \Pr_{\substack{\O=(\OracleA,\OracleB,\OracleC)\gets \D^{(n)}\\x\gets \Samp^\OracleA(1^n)}}[M^{\O}(x)=f_n^\OracleA(x)]}\\
            \leq & \sum_{i=n}^{T(n)} 2^i\cdot \frac{\poly(i)}{2^{i^2/2}}\leq \sum_{i=n}^{T(n)} \frac{\poly(n)}{2^{n^2/2-n}} = \frac{1}{2^{\Omega(n^2)}}\\
        \end{split}
    \end{equation}
\end{proof}

\begin{remark}
    Note that the same theorem applies for $M$ a randomized Turing machine by augmenting the output of $\Samp^\OracleA$ with random coins ignored by $f_n^\OracleA$.
\end{remark}

\subsection{Majority is average case hard}

We will show that the problem of salted majority is hard on average, even against non-uniform adversaries. In particular, let $\O = (\OracleA,\OracleB,\OracleC)$ be an oracle in the support of $\D$ with $\OracleA$ representing the multilinear extension of $\{\RO_n\}_{n\in \N}$. We will define the salted majority problem to be the following function $h^{\RO}:\{0,1\}^*\to \{0,1\}$ defined by
$$h^{\RO}(x) = 1 \iff |\{y \in \{0,1\}^{|x|} : \RO_{2|x|}(x,y) = 1\}| > 2^{|x|}/2$$
In particular, $h^{\RO}(x)$ is the majority problem on the function $\RO_{2|x|}(x,\cdot)$.

We will also consider the parallel repetition of the salted majority problem. In particular, define
$$H^{\RO}_r(x_1,\dots,x_r) = (h^{\RO}(x_1),\dots,h^{\RO}(x_r))$$
For $\O$ defined using $\RO$, we also write $h^\O:=h^{\RO}$ and $H_r^\O:=H_r^{\RO}$.

Note that for $\AC^0$ circuits taking as input just $\RO$, $H^{\RO}$ is very hard.

\begin{lemma}\label{lem:seededmajhard}
    There is a universal constant $\beta>0$ such that, for all fixed $d$ and polynomials $p,q$, the following holds. Let $r\leq p(n)$ and let $\mathcal C$ be a distribution over $\AC^0[\xor](2^{q(n)}, d)$ circuits. Then for all sufficiently large $n$,
    $$\Pr_{C\gets\mathcal C,\,\RO,\,x_1,\dots,x_r\gets \{0,1\}^n}[C(x_1,\dots,x_r,\RO) = H^{\RO}_r(x_1,\dots,x_r)] \leq e^{-\beta r}$$
\end{lemma}

\begin{proof}
    Let 
    $$\epsilon = \Pr_{C\gets\mathcal C,\,\RO,\,x_1,\dots,x_r\gets \{0,1\}^n}[C(x_1,\dots,x_r,\RO) = H^{\RO}_r(x_1,\dots,x_r)]$$
    and assume towards contradiction that $\epsilon > e^{-\beta r}$.

    We can transform $\mathcal C$ into a distribution over circuits $C'$ which solves $\MAJ^{\otimes \ceil{r/2}}$ with probability at least $\epsilon - (er/2^n)^{\ceil{r/2}}$ as follows.
    \begin{enumerate}
        \item On input $Y_1,\dots,Y_{\ceil{r/2}} \in \{0,1\}^{2^n}$
        \item Sample $C\gets\mathcal C$. Sample $x_1, \dots, x_r \in \{0,1\}^n$ uniformly at random
        \item If there are less than $\ceil{r/2}$ distinct values of $x_i$, reject.
        \item Let $\{x_1',\dots,x_{\ceil{r/2}}'\}$ be the first $\ceil{r/2}$ distinct values of $x_1,\dots,x_r$.
        \item For all $t\neq 2n$, sample $\RO_t$ uniformly at random.
        \item For all $x' \notin \{x_1',\dots,x_{\ceil{r/2}}'\}$, sample $\RO_{2n}(x',\cdot)$ uniformly at random.
        \item For all $i\in [\ceil{r/2}]$, set $\RO_{2n}(x_i',\cdot) = Y_i$
        \item Run $C$ on $(x_1,\dots,x_r,\RO)$ and output the response on $x_1',\dots,x_{\ceil{r/2}}'$.
    \end{enumerate}
    This can be implemented by an $\AC^0[\xor]$ circuit of the same size and depth as $C$.

    Let $m=\ceil{r/2}$ and let $K$ be the number of distinct sampled salts. We have
    \begin{equation*}
        \begin{aligned}
            &\Pr_{C',Y_1,\dots,Y_m}[C'(Y_1,\dots,Y_m)
                = (\MAJ(Y_1),\dots,\MAJ(Y_m))]\\
            &\qquad\geq \Pr_{C\gets\mathcal C,\,\RO,\,x_1,\dots,x_r}
                [C(x_1,\dots,x_r,\RO)=H_r^{\RO}(x_1,\dots,x_r)
                    \text{ and }K\geq m]\\
            &\qquad\geq \epsilon-\left(\frac{er}{2^n}\right)^m.
        \end{aligned}
    \end{equation*}
    where $\left(\frac{er}{2^n}\right)^{\ceil{r/2}}$ is an upper bound on the probability that there are fewer than $\ceil{r/2}$ distinct samples.

    Set $\delta=1/4$ and choose a universal $0<\beta<c\delta/2$,
    where $c$ is from~\Cref{thm:directproduct}. Since $r\leq p(n)$,
    the collision term is at most $e^{-\Omega(nm)}$. As $r\leq2m$,
    for all sufficiently large $n$ we have
    \[
        \epsilon-\left(\frac{er}{2^n}\right)^m
        >e^{-\beta r}-\left(\frac{er}{2^n}\right)^m
        \geq e^{-c\delta m}.
    \]
    Fix the randomness of $C'$ so that its success probability is at
    least this bound. Applying~\Cref{thm:directproduct} with $t=m$
    and input length $N=2^n$, then substituting $C'$ for its oracle gates,
    gives a constant-depth $\AC^0[\xor]$ circuit for majority with
    agreement at least $3/4$ and size at most
    \[
        S'=2^{q(n)}\cdot
        \poly(2^n,m,\log(1/\delta),e^{c\delta m})
        =2^{\poly(n)}.
    \]
    
    Since $\polylog(S')=\poly(n)$, by~\Cref{thm:majoritylb} every circuit
    of this size has agreement at most
    $\frac{1}{2}+\frac{\poly(n)}{2^{n/2}}<\frac{3}{4}$ for sufficiently
    large $n$. This contradicts the agreement of the reconstructed circuit.
\end{proof}

\begin{theorem}[Weak hardness of salted majority]\label{thm:weakhard}
    There exists a universal constant $\delta$ such that with probability $1$ over $\O\gets \D$ the following holds.
    Let $p(n)$ be a polynomial.  Let $\{C^{\O}_n\}_{n\in \N}$ be a family of circuits of size $p(n)$. Then, for all sufficiently large $n$,
    $$\Pr_{x\gets \{0,1\}^n}[C_n^\O(x) = h^\O(x)] \leq 1-\delta$$
\end{theorem}

\begin{proof}
    Fix some polynomial $p(n)$. Let $BAD^\O_{n,p,\delta}$ be the event (over randomness the choice in oracle) that for some circuit $C^\O$ of size $\leq p(n)$ we have
    $$\Pr_{x \gets \{0,1\}^n}[C^\O(x) = h^\O(x)] \geq 1-\delta$$
    By padding, we can assume without loss of generality that the size of $C^\O$ is $p(n)$. For simplicity, we will say that the number of circuits of size $p(n)$ is at most $2^{p(n)}$. In fact, this is true for a larger polynomial $p'(n)$, and the same proof will follow by replacing $p(n)$ with $p'(n)$. 
    
    It now suffices to show that there exists some constant $\delta$ such that for all polynomials $p$, the probability over $\O \gets \D$ that $BAD_{n,p,\delta}^\O$ occurs infinitely often is $0$.

    \paragraph{Removing the Forrelated part of the oracle}
    Sample a new oracle $\O' \gets \D^{(2n)}$.
    We will first show that for all sufficiently large $n$,
    \begin{equation}\label{eq:want-to-show}
        \Pr_{\O\gets \D}[BAD^\O_{n,p,\delta}] \leq \Pr_{\O\gets \D^{(2n)}}[{BAD}^\O_{n,p,3\delta}] + 2^{-\Omega(n^2)}
    \end{equation}
    In particular, let $\ell\in\N$ be values to be set later and define $\Test^\O$ to be the following Turing machine: "On input $(k_1,\dots,k_\ell,s_1,\dots,s_\ell)$, output $1$ if and only if there is an oracle circuit $C^\O$ of size $p(n)$ such that $C^\O(k_i) = s_i$ for at least $1-2\delta$ fraction of $i\in [\ell]$."

    Note that since $\P^\O = \NP^\O$, this Turing machine can run in time $\poly(n,\ell)$. Let $\Samp^\OracleA(1^n)$ be the distribution which samples $(k_1,\dots,k_\ell)$ uniform random distinct elements from $\{0,1\}^n$, and sets $s_i = h^\O(k_i)$.

    For an oracle $\O$ and an oracle circuit $C^{(\cdot)}$, define
    $$a(\O,C^{(\cdot)}) = \Pr_{k \gets \{0,1\}^n}[C^\O(k) = h^\O(k)]$$
    For a sequence $\vec{k} = (k_1,\dots,k_\ell)$, define
    $$\wt{a}(\O,C^{(\cdot)},\vec{k}) = \Pr_{i\gets [\ell]}[C^{\O}(k_i) = h^\O(k_i)]$$

    Hoeffding's inequality gives that for all oracles $\O$, circuits $C^{(\cdot)}$,
    $$\Pr_{(k_1,\dots,k_\ell)\text{ distinct}}[|a(\O,C^{(\cdot)}) - \wt{a}(\O,C^{(\cdot)},\vec{k})| > \delta] \leq 2e^{-2\ell \delta^2}$$
    Union bound then says that with probability $\geq 1-2^{p(n)+1}e^{-2\ell\delta^2}$ over distinct. random $(k_1,\dots,k_\ell)$, for all circuits $C^{(\cdot)}$ of size at most $p(n)$,
    $$|a(\O,C^{(\cdot)}) - \wt{a}(\O,C^{(\cdot)},\vec{k})| \leq \delta$$

    Let $\epsilon = 2^{p(n)+1}e^{-2\ell\delta^2}$. If $BAD_{n,p,\delta}^\O$ is true, then there exists a circuit such that $a(\O,C^{(\cdot)}) \geq 1 - \delta$. With probability $\geq 1-\epsilon$ over the choice of distinct $(k_1,\dots,k_\ell)$, for that $C^\O$ we have
    $$\wt{a}(\O,C^{(\cdot)},\vec{k}) \geq 1-2\delta$$
    Thus, $\Pr_{X\gets \Samp^\OracleA}[\Test^\O(X)\to 1]\geq 1-\epsilon$.

    Similarly, if $BAD_{n,p,3\delta}^\O$ is false, then for all circuits $C^{(\cdot)}$ of size at most $p(n)$, $a(\O,C^{(\cdot)}) < 1-3\delta$. So with probability $\geq 1-\epsilon$ over the choice of distinct $(k_1,\dots,k_\ell)$, for all circuits $C^\O$ we have
    $$\wt{a}(\O,C^{(\cdot)},\vec{k}) < 1-2\delta$$
    Thus, $\Pr_{X\gets \Samp^\OracleA}[\Test^\O(X)\to 1] \leq \epsilon$.

    In particular, we have for all distributions $\mathcal{R}$ over oracles,
    \[
        \begin{aligned}
            \Pr_{\O\gets\mathcal R}[BAD^\O_{n,p,\delta}]-\epsilon
            &\leq \Pr_{\O\gets\mathcal R,\,X\gets \Samp^\OracleA(1^n)}
                [\Test^\O(X)=1]\\
            &\leq \Pr_{\O\gets\mathcal R}[BAD^\O_{n,p,3\delta}]
                +\epsilon.
        \end{aligned}
    \]
    
    By~\Cref{lem:classicalforr} with parameter $2n$,
    \[
        \left|
            \Pr_{\substack{\O\gets\D\\X\gets \Samp^\OracleA(1^n)}}[\Test^\O(X)=1]
            -\Pr_{\substack{\O\gets\D^{(2n)}\\X\gets \Samp^\OracleA(1^n)}}[\Test^\O(X)=1]
        \right|\leq 2^{-\Omega(n^2)}
    \]
    Taking $\ell = \Omega(n^2p(n)/\delta^2)$ gives $\epsilon = 2^{-\Omega(n^2)}$. And so we get~\Cref{eq:want-to-show}.

    \paragraph{Bounding $\Pr_{\O\gets \D^{(2n)}}[BAD_{n,p,3\delta}]$}
    Let $\O$ be such that $BAD_{n,p,3\delta}^\O$ holds. Define a Turing machine $M$ solving $H^\O_r$ (for some $r$ to be set later) as follows
    \begin{enumerate}
        \item On input $x_1,\dots,x_r\in \{0,1\}^n$
        \item Pick a circuit $C^\O$ of size $p(n)$ uniformly at random
        \item For each $i\in [r]$, run $C^\O(x_i) \to b_i$
        \item Output $(b_1,\dots,b_r)$
    \end{enumerate}

    Note that if $BAD_{n,p,3\delta}^\O$ holds, then with probability at least $2^{-p(n)}$ $C^\O$ will solve salted majority with probability $1-3\delta$. Since the $x_i$s are sampled independently, we get
    $$\Pr_{x_1,\dots,x_r\gets \{0,1\}^n}[M^\O(x_1,\dots,x_r) = H^\O_r(x_1,\dots,x_r)] \geq 2^{-p(n)}\cdot (1-3\delta)^r$$

    And so in particular we get
    $$\Pr_{\substack{\O\gets \D^{(2n)}\\x_1,\dots,x_r\gets \{0,1\}^n}}[M^\O(x_1,\dots,x_r) = H^\O_r(x_1,\dots,x_r)] \geq 2^{-p(n)}\cdot (1-3\delta)^r \cdot \Pr_{\O\gets \D^{(2n)}}[BAD_{n,p,3\delta}^\O]$$

    Now, we apply~\Cref{cor:ac02} to get an $\AC^0[\xor]$ circuit solving salted majority. In particular, for $\O=(\OracleA,\OracleB,\OracleC) \gets \D^{(2n)}$, with $\OracleA$ the multilinear extension of $\RO$, there exists some constant $d$ (depending on $p(n)$) such that we have an $\AC^0[\xor](2^{1+n^{2^d}},2d+2)$ circuit $C'$ taking in input $$x_1,\dots,x_r,(\RO_1,\dots,\RO_{n^{2^{d-1}}}),(\OracleB_1,\dots,\OracleB_{n^{2^{d-1}}}),(\OracleC_1,\dots,\OracleC_n)$$ such that
    $$\Pr_{\substack{\O\gets \D^{(2n)}\\x_1,\dots,x_r\gets \{0,1\}^n}}[C'(x_1,\dots,x_r) = H^\O_r(x_1,\dots,x_r)] \geq 2^{-p(n)}\cdot (1-3\delta)^r \cdot \Pr_{\O\gets \D^{(2n)}}[BAD_{n,p,3\delta}^\O]$$

    But note that since $\O$ is drawn from $\D^{(2n)}$, the auxiliary input of $C'$ (corresponding to the desciption of the oracles $\OracleB$ and $\OracleC$) is independent of $\RO_{2n}$. And so by~\Cref{lem:seededmajhard} we get that for all sufficiently large $n$,
    $$2^{-p(n)}\cdot (1-3\delta)^r \cdot \Pr_{\O\gets \D^{(2n)}}[BAD_{n,p,3\delta}]\leq e^{-\beta r}$$

    Therefore,
    $$\Pr_{\O\gets \D^{(2n)}}[BAD_{n,p,3\delta}^\O] \leq 2^{p(n)}\cdot (1-3\delta)^{-r}\cdot e^{-\beta r}$$

    Setting $r=\lceil2n^2\cdot p(n)/\beta\rceil$ and $\delta$ such that $(1-3\delta) = e^{-\beta/2}$, we get
    \begin{equation*}
        \Pr_{\O\gets \D}[BAD_{n,p,\delta}^\O] \leq \Pr_{\O\gets \D^{(2n)}}[BAD_{n,p,3\delta}^\O] + 2^{-\Omega(n^2)} \leq 2^{p(n)}\cdot e^{-\beta r/2} + 2^{-\Omega(n^2)} = 2^{-\Omega(n^2)}
    \end{equation*}

    We then have
    $$\sum_{n=1}^{\infty} \Pr_{\O\gets \D}[BAD_{n,p,\delta}^\O] = \sum_{n=1}^{\infty} 2^{-\Omega(n^2)} < \infty$$
    and so by Borel-Cantelli, for all polynomials $p$, 
    $$\Pr_{\O \gets \D}[BAD_{n,p,\delta}^\O\text{ occurs infinitely often}] = 0$$

    Note that every polynomial $p$ is bounded by $(n+2)^c$ for some constant $c$. And so taking the union bound over the countable set of all such polynomials we get that with probability $1$ over $\O\gets \D$, for all polynomials $p$, $BAD_{n,p,\delta}^\O$ does not occur infinitely often.
\end{proof}

\subsection{Applications}

\paragraph*{Proofs of Quantumness}

\begin{corollary}
    With probability $1$ over $\O\gets \D$, there exists a public coin, efficiently verifiable proof of quantumness.
\end{corollary}

\begin{proof}
    Let $\delta$ be as in~\Cref{thm:weakhard}.

    Consider the following construction
    \begin{enumerate}
        \item Repeat the following trial $n$ times. If all trials pass, the verifier accepts, otherwise it rejects.
        \item In trial $i$, the verifier samples $k_i\gets \{0,1\}^n$ and sends $k_i$ to the prover.
        \item The prover responds with $z_i = h^{\O}(k_i)$.
        \item Then the prover and verifier engage in a public-coin interactive proof that $z_i = h^\O(k_i)$, with a $\BQP^\O$ honest prover, amplified to make its soundness error at most $\delta/2$ and its completeness error negligible.
        \item If the verifier accepts the proof, then the trial is passed. Otherwise, the verifier rejects.
    \end{enumerate}

    Correctness follows from the fact that $h^\O \in \PSPACE^{\RO} \subseteq \IP^{\OracleA}$. Note that all languages in $\IP^{\OracleA}$ have an interactive proof with a prover implementable in $\PSPACE^{\OracleA}$ that is public-coin, has negligible completeness error, and arbitrary constant soundness error~\cite{STOC:GolSip86}. Since $\PSPACE^{\OracleA} \subseteq \BQP^\O$, the honest prover can be implemented by a QPT oracle machine.

    By viewing the previous rounds and the prover's state as classical advice,~\Cref{thm:weakhard} gives that, conditioned on the previous trials passing, a classical prover finds $z_i = h^\O(k_i)$ with probability at most $1-\delta$. If $z_i\neq h^\O(k_i)$, the interactive proof accepts with probability at most $\delta/2$. Thus each trial passes with probability at most $1-\delta/2$, conditioned on the previous trials passing, and the probability that all trials pass is at most $(1-\delta/2)^n = \negl(n)$.
\end{proof}

\begin{theorem}[Stockmeyer's approximate counting {\cite{STOC:Stockmeyer83}}]\label{lem:stockmeyer-counting}
    Let $\O$ be a classical oracle such that $\P^\O=\NP^\O$, let $R^\O(x,y)$ be a deterministic polynomial-time oracle predicate, and let $m$ be a polynomial.
    Define
    \[
        N(x)=\bigl|\{y\in\{0,1\}^{m(|x|)}:R^\O(x,y)=1\}\bigr|.
    \]
    For any inverse-polynomial accuracy parameter $\alpha\in(0,1)$, there exists a deterministic polynomial-time algorithm with oracle access to $\O$ outputting an estimate $\widehat N(x)$ satisfying
    \[
        (1-\alpha)N(x)\leq\widehat N(x)\leq(1+\alpha)N(x)
    \]
    for all $x$. In particular, zero counts are recognized exactly.
\end{theorem}

The following corollary is folklore, but we provide the full proof for completeness.

\begin{corollary}[Classical provers for constant-round protocols]\label{thm:classical-prover-simulation}
    Let $\O$ be a classical oracle such that $\P^\O=\NP^\O$.
    Let $V^\O$ be a classical probabilistic polynomial-time verifier in a constant-round protocol with classical communication.
    Then there exists a deterministic polynomial-time classical prover $\ol{P}^\O$ such that, for every input $x$ and every prover $P^\O$ (including quantum and inefficient provers),
    \[
        \Pr[(V^\O\leftrightarrows\ol{P}^\O)(x)\to 1]
        \geq \Pr[(V^\O\leftrightarrows P^\O)(x)\to 1]-0.01.
    \]
\end{corollary}

\begin{proof}
    Let $c$ be the number of prover messages used in the protocol. Fix a polynomial bound $m(|x|)$ on the randomness of the verifier. Let $T_i$ be a partial transcript after $i$ prover messages and any following verifier message. Say that coins $r$ are consistent with $(x,T_i)$ if $V^\O(x,r)$ produces the verifier messages in $T_i$ when receiving the prover messages in $T_i$. We consider transcripts with at least one consistent choice of coins. Define
    $$\OPT(x,T_i) = \max_{P^\O}\Pr[V^\O(x,r)\leftrightarrows P^\O \to 1 | r\text{ consistent with }(x,T_i)]$$
    Here $T_i$ is fixed, and $P^\O$ chooses only the remaining messages.

    Set $\nu_i = \frac{1}{100\cdot 3^i}$. We will inductively define deterministic strategies $S_i(x,T_i)$ taking in a transcript for the first $i$ rounds and producing a strategy for the remaining. Let
    $p_i(x,T_i)$ be the probability over $r$ consistent with $(x,T_i)$ that a verifier receiving the prover messages in $T_i$ for the first $i$ rounds and $S_i(x,T_i)$ for the remaining rounds outputs $1$.
    We will prove that for all $i\in\{0,\dots,c\},x,T_i$,
    $$p_i(x,T_i)\geq \OPT(x,T_i) - \nu_i$$

    Define $S_c$ to do nothing. The inductive hypothesis holds since the prover sends no messages after round $c$.

    Now, for each partial transcript $T_i$ and for each message $a$ sent by the prover in round $i+1$, call $r$ "good" for $(x,T_i)$ if $(x,r)$ is consistent with $T_i$ and $V(x,r)$ accepts when receiving prover messages $T_i$, then $a$, then the messages produced by strategy $S_{i+1}$. Define
    $$N_{x,T_i}(a) = \abs{\{r \in \{0,1\}^{m(|x|)} | r\text{ is "good"}\}}$$
    Note that given $x,T_i$, we can check if $r$ is "good" in polynomial time by simply running the verifier with the corresponding strategy. Here, $S_{i+1}$ runs in polynomial time by the inductive hypothesis.

    Define $\wt{N}_{x,T_i}(a)$ be the estimate for $N_{x,T_i}(a)$ produced by~\Cref{lem:stockmeyer-counting} with parameter $\nu_{i+1}$. Since $\P^\O = \NP^\O$ and since $i+1 \leq c$ is a constant, this is computable in deterministic polynomial time and for all $a,x,T_i$,
    $$(1-\nu_{i+1})N_{x,T_i}(a) \leq \wt{N}_{x,T_i}(a) \leq (1+\nu_{i+1})N_{x,T_i}(a)$$

    Define ${a}^*$ be any value maximizing ${N}_{x,T_i}(a)$ and let $\wt{a}^*$ be any value maximizing $\wt{N}_{x,T_i}(a)$. Note that since $\P^\O = \NP^\O$ and $\wt{N}_{x,T_i}(a)$ is computable in polynomial time, $\wt{a}^*$ is also computable in polynomial time. Furthermore, we can bound $N_{x,T_i}(a^*)$ by $N_{x,T_i}(\wt{a}^*)$ as follows 
    \begin{equation*}
        \begin{split}
            N_{x,T_i}(a^*) &\leq \frac{1}{1-\nu_{i+1}} \wt{N}_{x,T_i}(a^*)\\
            &\leq \frac{1}{1-\nu_{i+1}} \wt{N}_{x,T_i}(\wt{a}^*)\\
            &\leq \frac{1+\nu_{i+1}}{1-\nu_{i+1}} {N}_{x,T_i}(\wt{a}^*)
        \end{split}
    \end{equation*}

    We then define $S_i$ to output in round $i+1$ message $\wt{a}^*$, and then run $S_{i+1}$ for the remaining rounds. By the inductive hypothesis, averaging over the next verifier message, $N_{x,T_i}(a^*)$ divided by the number of consistent coins is at least $\OPT(x,T_i)-\nu_{i+1}$. Thus,
    \begin{equation}
        \begin{split}
            p_i(x,T_i) &\geq \frac{1-\nu_{i+1}}{1+\nu_{i+1}}\bigl(\OPT(x,T_i)-\nu_{i+1}\bigr)\\
            &\geq \OPT(x,T_i) - 3\nu_{i+1}\\
            &= \OPT(x,T_i) - \nu_i
        \end{split}
    \end{equation}

    And so setting $\ol{P}^\O = S_0$ completes the proof.
\end{proof}

Thus, since with probability $1$ over $\O\gets \D$ we have $\P^\O = \NP^\O$, every constant-round proof of quantumness has an efficient classical prover and so therefore cannot be secure. 

\begin{corollary}
\label{coro:no_const_PoQ}
    With probability $1$ over $\O\gets \D$, there does not exist a constant-round proof of quantumness relative to $\O$.
\end{corollary}

Combining~\Cref{thm:directproduct,thm:weakhard} we get the following corollary.

\begin{corollary}
    With probability $1$ over $\O\gets \D$, there exists a two message quantumly-verifiable proof of quantumness.
\end{corollary}

\begin{proof}
    The verifier samples $k_1,\dots,k_n\gets \{0,1\}^n$ and sends them to the prover, who responds with $z_1,\dots,z_n$. The verifier accepts if and only if for all $i$, $z_i=h^\O(k_i)$. This can be computed efficiently since $h^\O \in \PSPACE^\O \subseteq \BQP^\O$. Repeat each computation enough times that the probability of any incorrect check is negligible.

    Suppose a polynomial-size classical prover answers all $n$ instances correctly with non-negligible probability. Then this probability is at least $1/q(n)$ for infinitely many $n$, for some polynomial $q(n)\geq 2$. Apply~\Cref{thm:directproduct} with $t=n$ and accuracy parameter $\eta=\ln(q(n))/(cn)$, where $c$ is the constant in that theorem. This gives a classical oracle circuit of polynomial size agreeing with $h^\O$ on at least a $1-\eta$ fraction of inputs. Since $\eta\to 0$, this contradicts~\Cref{thm:weakhard} for sufficiently large $n$. Thus the prover's success probability, including the verifier's error, is negligible.
\end{proof}

\section{A quantum oracle where $\BQP=\QCMA$}

\begin{theorem}
    There exists a distribution over unitary oracles $\O\gets \D$ such that with probability $1$,
    \begin{enumerate}
        \item $\BQP^{\O,\PSPACE} = \QCMA^{\O,\PSPACE}$
        \item Relative to $\O,\PSPACE$, quantum verifier oblivious state preparation exists
    \end{enumerate}
\end{theorem}

\subsection{The oracle}

We define a distribution $\D$ over unitary oracles $\O = (\SG,\Expand,\Mix)\gets \D$ consisting of three parts. We will have $\SG = \{\SG_t\}_{t\in \N}$, $\Expand = \{\Expand_t\}_{t\in \N}$, $\Mix = \{\Mix_t\}_{t\in \N}$. We will write $\O_t = (\SG_t,\Expand_t,\Mix_t)$.

For each $t$, let $\RO_t:\{0,1\}^{t}\times \{0,1\}^{4t} \to \{0,1\}^{6t}\times \{0,1\}^{6t}$ be a random oracle. Let $W_t^{\SG}:\mathcal{H}(\{0,1\}^{4t}) \to \mathcal{H}(\{0,1\}^{4t})$ be random unitaries. For each $x\in \{0,1\}^t$, we will also sample a random state $\ket{\phi_x}\gets \HaarSt(\{0,1\}^{4t})$ and a random unitary $W_{t,x}^{\Expand}:\mathcal{H}(\{0,1\}^{4t+1})\to \mathcal{H}(\{0,1\}^{4t+1})$.

Fix any $t\in \N$. We will define some important states used in the definition of our oracle.
\begin{enumerate}
    \item We denote the output of the random oracle on an input $x\in\{0,1\}^t$, $y\in\{0,1\}^{4t}$ by
    $$\RO_t(x,y) = (u_{x,y},v_{x,y})$$
    \item For $x\in \{0,1\}^t$, $y\in \{0,1\}^{4t}$, we define the claw state
    $$\ket{c_{x,y}} \propto \left(\ket{u_{x,y}} + \ket{v_{x,y}}\right)$$
    Note that the normalization differs depending on whether $u_{x,y} = v_{x,y}$.
    \item We define the random index mixed state $\sigma_t$ to sample a random key $x\in \{0,1\}^t$ and output $\ket{x}\ket{\phi_x}$. Formally,
    $$\sigma_t = \frac{1}{2^t}\sum_{x\in \{0,1\}^t} \ketbra{x}\otimes \ketbra{\phi_x}$$
    \item For $x\in \{0,1\}^t$, we define a mixed state used for preparing random claw states $\rho_x$ to sample a random key $y\in \{0,1\}^{4t}$ and output $\ket{y}\ket{c_{x,y}}$. Formally,
    $$\rho_x = \frac{1}{2^{4t}} \sum_{y\in \{0,1\}^{4t}} \ketbra{y}\otimes \ketbra{c_{x,y}}$$
\end{enumerate}

Now that we have these states defined, we can define the component oracles as follows.
\begin{enumerate}
    \item $\SG_t$ will be a state preparation unitary for $\sigma_t$ indexed by $W_t^{\SG}$. That is,
    $$\SG_t = S_{\sigma_t}^{W_t^{\SG}}$$
    \item $\Expand_t$ will take in an input $x$ and prepare $\rho_x$, with the corresponding state preparation unitary indexed by $W_{t,x}^{\Expand}$. That is,
    $$\Expand_t = \sum_{x\in \{0,1\}^t} \ketbra{x} \otimes S_{\rho_x}^{W_{t,x}^{\Expand}}$$
    \item $\Mix_t$ will take in inputs $x$, $y$, and $\ket{\phi_x}$ and prepare $\ket{\RO_t(x,y)} = \ket{u_{x,y},v_{x,y}}$. It will act as the identity on all orthogonal inputs. Formally,
    $$\Mix_t = \sum_{\substack{x\in\{0,1\}^t\\y\in\{0,1\}^{4t}}} \ketbra{x,y} \otimes \left(\ketbra{\phi_x} \otimes S_{\ket{u_{x,y},v_{x,y}}} + (I - \ketbra{\phi_x}) \otimes I\right)$$
\end{enumerate}

Independently of $\O$, fix an oracle $\PSPACE$ for a $\PSPACE$-complete problem, with $\PSPACE_t$ denoting its restriction to inputs of length $t$. We write superscripts $\O,\PSPACE$ when an algorithm has access to both. The simulations below also allow access to this fixed oracle, which is independent of all sampled states, unitaries, and random functions.

\subsection{Simulating the oracle with random oracle points}\label{sec:oraclesim}

We will show that with high probability over the oracle, any algorithm querying the oracle can be simulated by an adversary which just sees samples from the oracle. Notably, this will hold even for adversaries which get both classical advice on the random unitaries and states used by the oracle as well as the entire truth table of the classical random function $\RO$. 

In particular, for a given random oracle $\RO_n$ and a parameter $L(n)$, define the following mixed state $\aux_{\RO_n}:=\aux_{n,L(n)}^{\RO} = (\aux_{\SG},\aux_{\Expand},\aux_{\Mix})$ by
\begin{enumerate}
    \item Sample a list of $L(n)$ points $K=(k_1,\dots,k_{L(n)})\gets(\{0,1\}^n)^{L(n)}$. Let $K^\ast = \{k_i : i \in [L(n)]\}$ be the set of distinct points in $K$. Set $$\aux_{\SG} = \bigotimes_{i=1}^{L(n)}\ketbra{k_i}$$
    \item For each $x\in \{0,1\}^n$, sample $L(n)$ points $J_x = (y_{x,1},\dots,y_{x,L(n)})\gets(\{0,1\}^{4n})^{L(n)}$. Define the state $\ket{\alpha_x} = \bigotimes_{i \in [L(n)]} \ket{y_{x,i}}\ket{c_{x,y_{x,i}}}$. Define $$\aux_{\Expand} = \bigotimes_{x\in \{0,1\}^n} \ketbra{\alpha_x}$$
    \item For all $x\in \{0,1\}^n$, define $\ket{\beta_x} = \bigotimes_{y\in \{0,1\}^{4n}}\ket{\RO_n(x,y)}$. Set $$\aux_{\Mix} = \ketbra{K}\otimes \bigotimes_{k\in K^\ast} \ketbra{\beta_k}$$
\end{enumerate}
The mixed state $\aux_{\RO_n}$ is the average of $\aux_{\SG}\otimes\aux_{\Expand}\otimes\aux_{\Mix}$ over the joint sampling above. Its three registers may be correlated.

\begin{theorem}\label{thm:keylemma}
    Let $p(n),q(n)\geq 1$ be any computable functions such that $q(n)<2^{n/8}$ and $p(n)<2^{n/8}$. There exists a simulator $\Sim_{\aux}$ and a computable function $L(n)$ such that with probability $\geq 1-2^{-\Omega(n^2)}$ over the choice of oracle $\SG_n,\Mix_n,\Expand_n$, the following holds. Let $\A^{(\cdot),\PSPACE}$ be any oracle circuit of size at most $q(n)$ (and thus also making at most $q(n)$ queries to its oracle). Then, for all sufficiently large $n$,
    $$\abs{\Pr[\A^{\SG_n,\Expand_n,\Mix_n,\PSPACE}(\RO_n)\to 1] - \Pr[\A^{\Sim_{\aux}(\aux_{\RO_n}),\PSPACE}(\RO_n) \to 1]} \leq \frac{1}{p(n)}$$
    where $\RO_n$ corresponds to the entire truth table of the random function $\RO_n$ used by the oracle.

    Furthermore, we have that $L(n)<\poly(n,p(n),q(n))$ although $\Sim_{\aux}$ itself is inefficient. Note that by virtue of being a circuit, $\A$ implicitly gets non-uniform classical advice on the oracle.

\end{theorem}

\begin{proof}
    Since we will only consider level $n$ of the oracle $\O$, for the duration of this proof we will omit $n$ in relevant subscripts.

    The simulator will consist of three components $\Sim_{\SG},\Sim_{\Expand},\Sim_{\Mix}$ which will emulate the three components of $\O_n$. On initialization, the simulator will also split its advice into three. $\Sim_{\SG}$ will receive the state $\aux_{\SG}$, $\Sim_{\Expand}$ will receive the state $\aux_{\Expand}$, and $\Sim_{\Mix}$ will receive the state $\aux_{\Mix}$.

    Our simulator will be defined as follows:
    \begin{enumerate}
        \item $\Sim_{\Init}$: At initialization, for each distinct $k\in K^\ast$, the simulator will sample a Haar random state $\ket{r_k}\gets\HaarSt(\{0,1\}^{4n})$. It will provide classical descriptions of these states to the corresponding simulators\footnote{Note that a Haar random state does not have a finite classical description. This can be handled by replacing the Haar random states with a $t$-design for sufficiently large $t$.}. It will also split the registers of $\aux_{\RO_n}$ into its parts $(\aux_{\SG},\aux_{\Expand},\aux_{\Mix})$ and give each part to the corresponding simulator.
        \item $\Sim_{\SG}(\aux_{\SG})$: will parse $\aux_{\SG}$ as $K=(k_1,\dots,k_{L(n)})$. It will then run the simulator from~\Cref{lem:stateprep} on initial advice state $\bigotimes_{i=1}^{L(n)}\left(\ketbra{k_i}\otimes\ketbra{r_{k_i}}\right)$.
        \item $\Sim_{\Expand}(\aux_{\Expand})$: will run the simulator from~\Cref{cor:stateprep} on initial advice state $\aux_{\Expand}$.
        \item $\Sim_{\Mix}(\aux_{\Mix})$: will parse $\aux_{\Mix}$ as $K$ as well as $\{\RO_n(k,y)\}_{k\in K^\ast,y\in \{0,1\}^{4n}}$. It will then apply the map
        $$\Mix_K = I+\sum_{x\in K^\ast}\sum_{y\in\{0,1\}^{4n}}\ketbra{x,y}\otimes\ketbra{r_x}\otimes\left(S_{\ket{u_{x,y},v_{x,y}}}-I\right)$$
    \end{enumerate}

    Towards this end, we will define a few hybrid games. Each game $\Hyb_i$ will consist of an initialization phase $\Init^{(i)}$, followed by $\A^{(\cdot),\PSPACE}$ interacting with three (possibly stateful) oracles $\SG^{(i)},\Expand^{(i)},\Mix^{(i)}$ and outputting a single bit $b$, which will be the output of the game. They will be defined as follows:

    $\Hyb_0$: This will be the game where $\A$ interacts with $\O$ directly. That is,
    \begin{enumerate}
        \item $\Init^{(0)}$: does nothing
        \item $\SG^{(0)}$: queries $\SG$ on its input
        \item $\Expand^{(0)}$: queries $\Expand$ on its input
        \item $\Mix^{(0)}$: queries $\Mix$ on its input.
    \end{enumerate}

    $\Hyb_1$: This will be the same as $\Hyb_0$, except the unitaries and states used by the oracle will be replaced with fresh random unitaries and states, keeping $\RO_n$ fixed. That is,
    \begin{enumerate}
        \item $\Init^{(1)}$: Samples $\wt{W}^{\SG}\gets\Haar(\{0,1\}^{4n}),\quad \{\wt{W}_x^{\Expand}\}_{x\in\{0,1\}^n}\gets\Haar(\{0,1\}^{4n+1})$. For each $x$ samples a Haar random state $\ket{\wt{\phi}_x}\gets\HaarSt(\{0,1\}^{4n})$. Define
        $$\wt{\sigma} = \frac{1}{2^n} \sum_{x\in \{0,1\}^n} \ketbra{x}\otimes \ketbra{\wt{\phi}_x}$$
        \item $\SG^{(1)}$: applies the unitary
        $$S_{\wt{\sigma}}^{\wt{W}^{\SG}}$$
        \item $\Expand^{(1)}$: applies the unitary
        $$\sum_{x\in \{0,1\}^n} \ketbra{x} \otimes S_{\rho_x}^{\wt{W}_x^{\Expand}}$$
        \item $\Mix^{(1)}$: applies $\Mix$ with $\ket{\phi_x}$ replaced by $\ket{\wt{\phi}_x}$. That is, $\Mix^{(1)}$ applies the map
        $$\sum_{x\in \{0,1\}^n}\sum_{y\in \{0,1\}^{4n}} \ketbra{x,y} \otimes \left(\ketbra{\wt{\phi}_x} \otimes S_{\ket{u_{x,y},v_{x,y}}} + (I - \ketbra{\wt{\phi}_x}) \otimes I\right)$$
    \end{enumerate}

    The only difference between $\Hyb_0$ and $\Hyb_1$ is the replacement of the oracle's component unitaries by freshly random ones as well as the random states $\ket{\phi_x}$ by fresh random ones. We will apply~\Cref{thm:unitarysim} individually to each state $\ket{\phi_x}$ (viewed as a random unitary $U_k$ applied to $\ket{0}$) as well as each of $W^{\SG},W_x^{\Expand}$. In particular, we will set $\epsilon = 2^{-3n/2}$ for each replacement. We then get that for any given adversary $\A$, with probability
    $$\geq 1 - 2(2\cdot 2^n + 1)\cdot\exp\left(-\frac{(2^{4n}-2)2^{-3n}}{96(4q(n))^2}\right)$$
    $$\abs{\Pr[\Hyb_0\to 1] - \Pr[\Hyb_1\to 1]} \leq (2\cdot 2^n + 1)\cdot 2^{-3n/2}$$
    Note that giving $\A$ the description of $\RO_n$ does not change the adversaries query efficiency for the unitaries we replace nor its description length, and so~\Cref{thm:unitarysim} still applies.

    Taking the union bound over all adversaries $\A$, we get for all sufficiently large $n$, with probability $\geq 1-2^{O(q(n)(n+\log(q(n)+1)))}\cdot\exp(-2^{2n/3}) = 1-2^{-\Omega(n^2)}$, we have
    $$\abs{\Pr[\Hyb_0\to 1] - \Pr[\Hyb_1\to 1]} \leq 2^{-n/3}$$
    
    $\Hyb_2$: This will be the map which replaces $\SG^{(1)}$ and $\Expand^{(1)}$ with the simulators given by~\Cref{lem:stateprep,cor:stateprep}.
    \begin{enumerate}
        \item $\Init^{(2)}$:
        \begin{enumerate}
            \item Samples $\ket{\wt{\phi}_x} \gets \HaarSt(\{0,1\}^{4n})$ and
        $$\wt{\sigma} = \frac{1}{2^n} \sum_{x\in \{0,1\}^n} \ketbra{x}\otimes \ketbra{\wt{\phi}_x}$$
            \item Samples $k_1,\dots,k_{L(n)}\gets \{0,1\}^n$ uniformly at random.
            \item For each $x\in \{0,1\}^n$, samples $y_{x,1},\dots,y_{x,L(n)}\gets \{0,1\}^{4n}$ uniformly at random.
        \end{enumerate}
        \item $\SG^{(2)}$: Let $\Sim_{\SG}$ be the simulator from~\Cref{lem:stateprep} applied to $S_{\wt{\sigma}}^{\wt{W}^{\SG}}$. $\SG^{(2)}$ will run $\Sim_{\SG}$ with initial advice state $$\bigotimes_{i=1}^{L(n)}\left(\ketbra{k_i}\otimes\ketbra{\wt{\phi}_{k_i}}\right)$$
        \item $\Expand^{(2)}$: Let $\Sim_{\Expand}$ be the simulator from~\Cref{cor:stateprep} applied to
        $$\sum_{x\in \{0,1\}^n} \ketbra{x} \otimes S_{\rho_x}^{\wt{W}_x^{\Expand}}$$
        $\Expand^{(2)}$ will run $\Sim_{\Expand}$ with initial advice
        $$\bigotimes_{x\in\{0,1\}^n}\bigotimes_{i=1}^{L(n)}\left(\ketbra{y_{x,i}}\otimes\ketbra{c_{x,y_{x,i}}}\right)$$
        \item $\Mix^{(2)}$: this will be the same as $\Mix^{(1)}$.
    \end{enumerate}

    By instantiating~\Cref{lem:stateprep,cor:stateprep} with $\epsilon = \frac{1}{16p(n)},\quad L(n)=\ceil{16c\,p(n)q(n)^2}$, we have that for all choices of oracle $\O$ in the support of $\D$, for all adversaries $\A$,
    $$\abs{\Pr[\Hyb_2\to 1] - \Pr[\Hyb_1\to 1]} \leq \frac{1}{4p(n)}$$

    $\Hyb_3$: this will be the same as $\Hyb_2$, but $\Mix^{(2)}$ ignores all $x\notin K^\ast$ (i.e. acts the same way as $\Sim_{\Mix}$). Formally,
    \begin{enumerate}
        \item $\Init^{(3)}$: This will be the same as $\Init^{(2)}$. Let $K=(k_1,\dots,k_{L(n)})$ be the list of random points sampled.
        \item $\SG^{(3)}$: This will be the same as $\SG^{(2)}$
        \item $\Expand^{(3)}$: This will be the same as $\Expand^{(2)}$
        \item $\Mix^{(3)}$: This will apply the map
        $$I+\sum_{x\in K^\ast}\sum_{y\in\{0,1\}^{4n}}\ketbra{x,y}\otimes\ketbra{\wt{\phi}_x}\otimes\left(S_{\ket{u_{x,y},v_{x,y}}}-I\right)$$
    \end{enumerate}

    Fix $K$, the states $\ket{\wt{\phi}_k}$ for $k\in K^\ast$, and all randomness besides the remaining states $\ket{\wt{\phi}_x}$. Define the distribution over projectors $\Pi_{bad} = \sum_{x\in\{0,1\}^n\setminus K^\ast}\ketbra{x}\otimes\ketbra{\wt{\phi}_x}$. Note that $\Mix^{(3)}$ is now fixed and $\Mix^{(3)} (I - \Pi_{bad}) = \Mix^{(2)} (I - \Pi_{bad})$, and so by~\Cref{lem:reprogramming} it is sufficient to bound $\sup_{\ket{\phi}}\E[\norm{\Pi_{bad}\ket{\phi}}^2]$ where the expectation is over the remaining states $\ket{\wt{\phi}_x}$. But note that $\ket{\wt{\phi}_x}$ is Haar random for each $x\notin K^\ast$, and so
    \begin{equation*}
        \begin{split}
            \sup_{\ket{\phi}}\E[\norm{\Pi_{bad}\ket{\phi}}^2]\\
            \leq 2^n \sup_{\ket{\phi}}\E_{\ket{\psi}\gets \HaarSt(\{0,1\}^{4n})}[\abs{\bra{\phi}\ket{\psi}}^2] \\
            \leq 2^n\cdot 2^{-4n} \leq 2^{-3n}
        \end{split}
    \end{equation*}
    Applying~\Cref{lem:reprogramming}, we have that for all choices of oracle $\O$ in the support of $\D$, for all adversaries $\A$,
    $$\abs{\Pr[\Hyb_2\to 1] - \Pr[\Hyb_3\to 1]} \leq 2q(n)\sqrt{2^{-3n}}\leq \frac{1}{4p(n)}$$
    for sufficiently large $n$.

    $\Hyb_f$: This will be the game where the adversary interacts directly with the simulator. That is,
    \begin{enumerate}
        \item $\Init^{(f)}$: Uses the mixed state $\aux_{\RO_n}$ defined above. Runs $\Sim_{\Init}$ to generate $\{\ket{r_k}\}_{k\in K^\ast}$.
        \item $\SG^{(f)}$: Runs $\Sim_{\SG}(\aux_{\SG})$.
        \item $\Expand^{(f)}$: Runs $\Sim_{\Expand}(\aux_{\Expand})$.
        \item $\Mix^{(f)}$: Runs $\Sim_{\Mix}(\aux_{\Mix})$.
    \end{enumerate}

    Note that the only difference between $\Hyb_3$ and $\Hyb_f$ is syntactic, they compute the exact same functionality. And so $\Pr[\Hyb_3\to 1] = \Pr[\Hyb_f\to 1]$.

    Since $p(n)<2^{n/8}$, putting this all together we get that with probability $1-2^{-\Omega(n^2)}$, for all oracle circuits $\A$ of size at most $q(n)$,
    $$\abs{\Pr[\Hyb_0\to 1] - \Pr[\Hyb_f\to 1]} \leq \frac{1}{p(n)}$$

\end{proof}

\begin{corollary}\label{cor:auxrandom}
    Let $p(n),q(n)\geq 1$ be any computable functions such that $q(n)<2^{n/16+3}$ and $p(n)<2^{n/16+3}$. There exists a simulator $\Sim_{\aux}$ and a computable function $L(n)<\poly(n,\allowbreak p(n),\allowbreak q(n))$ such that with probability $\geq 1-2^{-\Omega(n)}$ over the choice of oracle $\SG_n,\Mix_n,\Expand_n$, the following holds. Let $\A^{(\cdot),\PSPACE}$ be any oracle circuit of size at most $q(n)$ (and thus also making at most $q(n)$ queries to its oracle). Then, for all sufficiently large $n$,
    $$\abs{\Pr[\A^{\SG_n,\Expand_n,\Mix_n,\PSPACE}\to 1] - \Pr_{\wt{\RO}_n}[\A^{\Sim_{\aux}(\aux_{\wt{\RO}_n}),\PSPACE} \to 1]} \leq\frac{1}{p(n)}$$
    where $\wt{\RO}_n$ is a fresh random oracle.
\end{corollary}

\begin{proof}
    Let $L(n)$ be as in~\Cref{thm:keylemma} with accuracy parameter $4p(n)$ and circuit-size bound $2q(n)+n$.

    In particular, define $\wt{\aux}_{\RO_n,\wt{\RO}_n}$ to be the same as $\aux_{\RO_n}$ where for each $i\in [L(n)]$, $\RO_n(k_i,\cdot)$ is replaced with $\wt{\RO}_n(k_i,\cdot)$. Whenever there is an overlap with a sampled pair $(x,y_{x,j})$, use $\RO_n(x,y_{x,j})$ in both places in $\wt{\aux}_{\RO_n,\wt{\RO}_n}$.

    Condition on the sampled $y$'s and the oracle values retained there, then apply~\Cref{cor:indis} to the larger function $x\mapsto \RO_n(x,\cdot)$ with these entries omitted. We get that for all algorithms $\A$ and all advice functions $c$ of output length $q(n)$,
    \begin{align*}
        &\abs{\Pr_{\RO_n}[\A^{\PSPACE}(c(\RO_n),\aux_{\RO_n}) \to 1]-\Pr_{\RO_n,\wt{\RO}_n}[\A^{\PSPACE}(c(\RO_n),\wt{\aux}_{\RO_n,\wt{\RO}_n}) \to 1]}\\
        &\qquad\leq \frac{(q(n)+n+1)L(n)}{2^{n/2}} + 2^{-n} \leq 2^{-n/6}
    \end{align*}
    for sufficiently large $n$. For $K=(k_1,\dots,k_{L(n)})$, the marginal distribution $k_i$ has full entropy, and so $H_{\infty}(k_i) = n$.

    We can then apply~\Cref{cor:indis} to the remaining $2^n\cdot L(n)$ sampled points $\RO_n(x,y_{x,i})$ to get
    \begin{align*}
        &\abs{\Pr_{\RO_n,\wt{\RO}_n}[\A^{\PSPACE}(c(\RO_n),\aux_{\wt{\RO}_n}) \to 1]-\Pr_{\RO_n,\wt{\RO}_n}[\A^{\PSPACE}(c(\RO_n),\wt{\aux}_{\RO_n,\wt{\RO}_n}) \to 1]}\\
        &\qquad\leq \frac{(q(n)+5n+1)\cdot 2^n L(n)}{2^{2n}} + 2^{-5n} \leq 2^{-n/2}
    \end{align*}
    for sufficiently large $n$. In particular, each $y_{x,i}$ has full min-entropy $4n$.

    Let
    $$\Delta(\RO_n)=\max_{\A}\abs{\Pr[\A^{\Sim_{\aux}(\aux_{\RO_n}),\PSPACE}\to 1]-\Pr_{\wt{\RO}_n}[\A^{\Sim_{\aux}(\aux_{\wt{\RO}_n}),\PSPACE}\to 1]}$$
    where the maximum is over circuits of size at most $q(n)$. Choose $c(\RO_n)$ to encode a maximizing circuit and the sign of the gap. By our size convention, this uses at most $q(n)$ bits. The two bounds above give $\E[\Delta(\RO_n)]\leq 2^{-n/6}+2^{-n/2}$, so Markov's inequality gives
    $$\Pr\left[\Delta(\RO_n)>\frac{3}{4p(n)}\right]\leq \frac{4p(n)}{3}\left(2^{-n/6}+2^{-n/2}\right).$$
    Combining this with the error $1/(4p(n))$ and failure probability $2^{-\Omega(n^2)}$ from~\Cref{thm:keylemma}, we get, for sufficiently large $n$, with probability at least $1-2^{-n/10}$ over the choice of $\SG_n,\Mix_n,\Expand_n$, for all oracle circuits $\A^{(\cdot),\PSPACE}$ of size at most $q(n)$,
    $$\abs{\Pr[\A^{\SG_n,\Expand_n,\Mix_n,\PSPACE}\to 1]-\Pr_{\wt{\RO}_n}[\A^{\Sim_{\aux}(\aux_{\wt{\RO}_n}),\PSPACE}\to 1]}\leq\frac{1}{p(n)}$$
\end{proof}

\begin{corollary}\label{cor:truerand}
    Let $p(n),q(n)\geq 1$ be any computable functions bounded above by $2^{n/16}$. With probability $\geq 1-2^{-n/10}$ over the choice of oracle $\SG_n,\Mix_n,\Expand_n$, the following holds. Let $\A^{(\cdot),\PSPACE}$ be any oracle circuit of size at most $q(n)$ (and thus also making at most $q(n)$ queries to its oracle). Then, for all sufficiently large $n$,
    $$\abs{\Pr[\A^{\SG_n,\Expand_n,\Mix_n,\PSPACE}\to 1] - \Pr_{\wt{\SG}_n,\wt{\Expand}_n,\wt{\Mix}_n}[\A^{\wt{\SG}_n,\wt{\Expand}_n,\wt{\Mix}_n,\PSPACE} \to 1]} \leq\frac{1}{2p(n)}$$
    where $\wt{\SG}_n,\wt{\Expand}_n,\wt{\Mix}_n$ is a freshly sampled set of oracles, including a fresh random function $\wt{\RO}_n$.
\end{corollary}

\begin{proof}
    The key idea is that we apply~\Cref{cor:auxrandom} twice. First for $\SG_n,\Expand_n,\Mix_n$ and second for a freshly random sampled $\wt{\SG}_n,\wt{\Expand}_n,\wt{\Mix}_n$. In particular, setting the error parameter as $\frac{1}{6p(n)}$, we get that with probability $\geq 1-2^{-n/10}$ over the choice of $\SG_n,\Expand_n,\Mix_n$,
    $$\begin{aligned}
        &\abs{\Pr[\A^{\SG_n,\Expand_n,\Mix_n,\PSPACE} \to 1] - \Pr_{\wt{\SG}_n,\wt{\Expand}_n,\wt{\Mix}_n}[\A^{\wt{\SG}_n,\wt{\Expand}_n,\wt{\Mix}_n,\PSPACE} \to 1]}\\
        &\qquad\leq\frac{1}{3p(n)}+2^{-n/10}\leq\frac{1}{2p(n)}
    \end{aligned}$$
    for all sufficiently large $n$. Here the extra $2^{-n/10}$ term comes from the probability that $\wt{\SG}_n,\allowbreak\wt{\Expand}_{n},\allowbreak\wt{\Mix}_n$ fails to satisfy the conditions of~\Cref{cor:auxrandom}. Both applications are within its parameter range since $6p(n)<2^{n/16+3}$.
\end{proof}

\begin{corollary}\label{cor:keylem}
    Let $q(n),p(n)\geq 1$ be any computable functions bounded above by $2^{n/16}$. There exists a simulator $\Sim_{\mathrm{eff}}$ such that with probability $\geq 1-2^{-n/10}$ over the choice of oracle $\SG_n,\allowbreak\Mix_n,\allowbreak\Expand_n$, the following holds. Let $\A^{(\cdot),\PSPACE}$ be any oracle circuit of size at most $q(n)$ (and thus also making at most $q(n)$ queries to its oracle). Then, for all sufficiently large $n$,
    $$\abs{\Pr[\A^{\SG_n,\Expand_n,\Mix_n,\PSPACE} \to 1] - \Pr[\A^{\Sim_{\mathrm{eff}},\PSPACE} \to 1]} \leq \frac{1}{p(n)}$$
    The simulator runs in time $\poly(n,p(n),q(n))$. Again, as $\A^{(\cdot),\PSPACE}$ is an oracle circuit, it implicitly gets advice about the oracle.
\end{corollary}

\begin{proof}
    Applying~\Cref{cor:truerand}, our simulator will implement $\wt{\SG}_n,\allowbreak\wt{\Expand}_n,\allowbreak\wt{\Mix}_n$ for a freshly random choice of random oracle, random states, and random unitaries. Note that each of $\wt{\SG}_n,\allowbreak\wt{\Expand}_n,\allowbreak\wt{\Mix}_n$ can be implemented efficiently by making polynomially many queries to fresh random oracles and unitaries. Furthermore, fresh classical random oracles and unitaries can be simulated efficiently using compressed oracles~\cite{C:Zhandry19} and path recording~\cite{STOC:MaHua25,GY26,FLMNW26} up to error $2^{-n}$ in time $\poly(n,p(n),q(n))$. The total error is at most $1/(2p(n))+2^{-n}\leq 1/p(n)$ for sufficiently large $n$.
\end{proof}

\subsection{Showing $\BQP^{\O,\PSPACE}=\QCMA^{\O,\PSPACE}$}

\begin{theorem}\label{thm:bqp-qcma-collapse}
    With probability $1$ over $\O\gets \D$, $\BQP^{\O,\PSPACE} = \QCMA^{\O,\PSPACE}$.
\end{theorem}

\begin{proof}
    Let $\Pi$ be any language in $\QCMA^{\O,\PSPACE}$ with verifier $\Ver^{\O,\PSPACE}$. For $n\in \N$, let $q(n)$ be a polynomial upper bound on the runtime of $\Ver^{\O,\PSPACE}(x,w)$ and on its circuit size after hardcoding $x,w$, for $x\in \{0,1\}^n,w\in \{0,1\}^{\poly(n)}$. This is thus also a bound on the number of queries made as well as the length of the witness. We will describe a QPT algorithm $\A^{\O,\PSPACE}$ which decides $\Pi$ as follows:
    \begin{enumerate}
        \item On input $x\in \{0,1\}^n$, let $d = \ceil{c\log n}$ for some appropriate constant $c$.
        \item Construct a new verifier $\wt{\Ver}^{\PSPACE}$ as follows
        \begin{enumerate}
            \item For $\ell\in [1,d]$, learn $(\SG_\ell,\Expand_\ell,\Mix_\ell)$ using~\Cref{thm:tomography} with error parameter $\frac{1}{3q(n)^3}$ and failure probability $2^{-d}/3$ for each component. This takes $2^{O(d)}\poly(q(n),d)=\poly(n)$ time, and all three estimates are accurate with probability at least $1-2^{-d}$. Define the corresponding oracles as $\wt{\O}_{\ell} = (\wt{\SG}_\ell,\wt{\Expand}_\ell,\wt{\Mix}_\ell)$.
            \item For $\ell \in [d+1,q(|x|)]$, let $\Sim_{\mathrm{eff},\ell}$ be as in~\Cref{cor:keylem} with $n=\ell$, circuit-size bound $Q_n=\max(q(n),2^{d/32})$, and accuracy parameter $P_n=\max(q(n)^2,2^{d/32})$. Set $\wt{\O}_\ell=\Sim_{\mathrm{eff},\ell}$. Since $P_n$ and $Q_n$ are both bounded above by polynomials in $n$, $\Sim_{\mathrm{eff},\ell}$ can be implemented efficiently.
            \item $\wt{\Ver}^{\PSPACE}$ will then run $\Ver^{\{\wt{\O}_\ell\}_{\ell\in [1,q(|x|)]},\PSPACE}$
        \end{enumerate}
        \item Finally, output $1$ if and only if there exists some witness $w$ such that $\Pr[\wt{\Ver}^{\PSPACE}(x,w)=1]\allowbreak \geq 0.5$, This can be computed using the $\PSPACE$ oracle since $\wt{\Ver}^{\PSPACE}$ is a polynomial time $\PSPACE$ oracle algorithm.
    \end{enumerate}

    To see that this succeeds with high probability over the oracle, it is sufficient to show that with high probability over the oracle and the randomness of $\A$, $\wt{\Ver}^{\PSPACE}$ agrees with $\Ver^{\O,\PSPACE}$ on all $x\in \{0,1\}^n$ and witnesses $w\in \{0,1\}^{q(|x|)}$. In particular, we will hybrid over all $\ell\in [q(n)]$. Define $\O^{(\ell)} = (\O_1,\dots,\O_\ell,\wt{\O}_{\ell+1},\dots,\wt{\O}_{q(|x|)})$.

    The following bounds hold simultaneously for all $x\in\{0,1\}^n$ and witnesses $w$.

    For all $\ell\in [1,d]$ and all $\O$ in the support of $\D$, with probability at least $1-2^{-d}$ over the coins of $\A$, for all $x\in\{0,1\}^n,w\in\{0,1\}^{q(n)}$,
    $$\abs{\Pr[\Ver^{\O^{(\ell)},\PSPACE}(x,w)=1] - \Pr[\Ver^{\O^{(\ell-1)},\PSPACE}(x,w)=1]} \leq \frac{1}{q(|x|)^2}$$

    For all $\ell \in [d+1,q(n)]$, as~\Cref{cor:keylem} relativizes we can apply it individually to each index. By~\Cref{cor:keylem} we have that for all sufficiently large $n$, with probability greater than $1-2^{-d/10}$, for all $x\in \{0,1\}^n,w\in \{0,1\}^{q(n)}$
    $$\abs{\Pr[\Ver^{{\O}^{(\ell)},\PSPACE}(x,w)\to 1] - \Pr[\Ver^{\O^{(\ell-1)},\PSPACE}(x,w)\to 1]} \leq \frac{1}{q(n)^2}$$
    Note that in particular since the bound holds for all circuits simultaneously, it holds for the circuits which hardcode $x$ and $w$ for all possible values of $x$ and $w$ simultaneously.

    And so by repeating all these hybrids, we get that for all sufficiently large $n$, with probability $\geq 1-d\cdot 2^{-d}-q(n)\cdot \left(2^{-d/10}\right)$
    $$\abs{\Pr[\Ver^{\O^{(q(n))},\PSPACE}(x,w)\to 1] - \Pr[\Ver^{\O^{(0)},\PSPACE}(x,w)\to 1]} \leq q(n)\cdot \frac{1}{q(n)^2} \leq 0.01$$

    By taking $c$ from $d=\ceil{c\log(n)}$ sufficiently large, the probability over $\O$ that one of the hybrids with $\ell>d$ fails for some $x,w$ is at most $q(n)2^{-d/10}\leq 1/n^2$. For every other oracle, tomography succeeds with probability at least $1-d2^{-d}\geq 2/3$, and whenever it succeeds the algorithm correctly decides all inputs of length $n$.

    Define $BAD_n$ to be the event over $\O$ that there exists an $x\in\{0,1\}^n$ on which $\A^{\O,\PSPACE}$ gives the wrong answer with probability greater than $1/3$.
    We have just shown that 
    $$\Pr[BAD_n]\leq \frac{1}{n^2}$$

    And so by Borel-Cantelli, the probability that $BAD_n$ occurs infinitely often is $0$. Thus, $\A^{\O,\PSPACE}$ correctly decides $\Pi$ with probability at least $2/3$ on all sufficiently large $n$, and so by hardcoding whether $x\in \Pi$ for all $x\in \{0,1\}^{n'}$ for $n'$ below this threshold, we get an algorithm which correctly decides $\Pi$ on all inputs with probability at least $2/3$. Thus, $\Pi\in \BQP^{\O,\PSPACE}$.

    Since the number of verifiers is countable, taking the union bound gives that with probability $1$ over the choice of oracle, $\QCMA^{\O,\PSPACE} \subseteq \BQP^{\O,\PSPACE}$.
\end{proof}

\subsection{Building oblivious state preparation}

We construct a claw state generator as in~\Cref{def:claw-state-generator} relative to our oracle.

\begin{lemma}[Decoding bound {\cite[Theorem~2.4]{FOCS:Nayak99}}]\label{lem:decoding-bound}
    Let $Z$ be uniform on $[M]$, and let $\{\rho_z\}_{z\in[M]}$ be density operators on a $d$-dimensional register.
    For any quantum decoding algorithm $\Dec$,
    \[
        \Pr[\Dec(\rho_Z)=Z]\leq \frac{d}{M}.
    \]
    The same bound holds if the decoder receives $\rho_Z\otimes\sigma$ for any fixed auxiliary state $\sigma$.
    In particular, a uniformly random $2n$-bit string encoded into $q$ qubits can be recovered with probability at most $2^{q-2n}$; for $q=n$, this is $2^{-n}$.
\end{lemma}

\begin{lemma}[Claw hardness]\label{lem:claw-hardness}
    Let $\A$ be any quantum algorithm, and let $m,\ell\geq 1$ be integers. Consider the following game:
    \begin{enumerate}
        \item Sample $a_1,\dots,a_\ell,b_1,\dots,b_\ell\gets \{0,1\}^m$ independently and uniformly at random.
        \item For each $i\in[\ell]$, define the normalized $m$-qubit claw state
        \[
            \ket{\chi_i}=\frac{\ket{a_i}+\ket{b_i}}{\sqrt{2+2\mathbf{1}[a_i=b_i]}}.
        \]
        \item Run $\A$ on $\bigotimes_{i=1}^{\ell}\ket{\chi_i}$ and output a pair $(a',b')$.
        \item Output $1$ if $(a',b') = (a_i,b_i)$ for some $i\in[\ell]$.
    \end{enumerate}
    Then the probability that $\A$ wins this game is at most $\ell\cdot 2^{-m}$. The same bound holds if $\A$ also receives any fixed auxiliary state independent of all sampled pairs.
\end{lemma}

\begin{proof}
    For each $i\in[\ell]$, let $E_i$ be the event that $\A$ outputs $(a_i,b_i)$.
    Fix $i$ and condition on arbitrary values of all pairs $(a_j,b_j)$ with $j\ne i$.
    The pair $(a_i,b_i)$ remains uniform on $\{0,1\}^{2m}$, and all other claw states, together with any additional fixed auxiliary state, form a fixed auxiliary state in tensor product with $\ket{\chi_i}$.
    Thus, $\A$ is a decoder for $(a_i,b_i)$ from the $m$-qubit encoding $\ket{\chi_i}$, together with independent auxiliary input.
    Applying~\Cref{lem:decoding-bound} with $M=2^{2m}$ and $d=2^m$ gives
    \[
        \Pr\!\left[E_i\,\middle|\,(a_j,b_j)_{j\ne i}\right]
        \leq \frac{2^m}{2^{2m}}
        =2^{-m}.
    \]
    Averaging over the conditioned pairs yields the same bound on $\Pr[E_i]$.
    Finally, by the union bound,
    \[
        \Pr[\A\text{ wins}]
        =\Pr\!\left[\bigcup_{i=1}^{\ell}E_i\right]
        \leq\sum_{i=1}^{\ell}\Pr[E_i]
        \leq \ell\cdot 2^{-m}.
    \]
\end{proof}

We now construct the claw state generator. Here $x,y$ index oracle inputs; the corresponding claw endpoints are $u_{x,y},v_{x,y}$. This will be a two message protocol as follows
\begin{construction}\label{cons:claw}\leavevmode
\begin{enumerate}
    \item The verifier begins by running $\SG_n$, measuring the register containing $\ket{x}$ and tracing out the purification and flag registers, to produce a state $\ket{x}\ket{\phi_x}$. The verifier sends $x$ to the prover.
    \item The prover runs $\Expand_n \ket{x}\ket{0}$ and traces out the purification and flag registers to produce the state $\ket{x}\ket{y}\ket{c_{x,y}}$. The prover sends $y$ to the verifier and outputs the claw-state $\ket{c_{x,y}}$.
    \item The verifier runs $\Mix_n \ket{x}\ket{y}\ket{\phi_x}\ket{0}$ and traces out the excess registers to produce the state $\ket{x}\ket{y}\allowbreak\ket{u_{x,y}}\allowbreak\ket{v_{x,y}}$. It then outputs $\ket{u_{x,y}}\ket{v_{x,y}}$.
\end{enumerate}
\end{construction}

\begin{claim}
    With probability $1$ over $\O\gets \D$,~\Cref{cons:claw} satisfies correctness.
\end{claim}

\begin{proof}
    This is self-evident by construction. That is, the only way correctness can be violated is if there exists some pair $x\in \{0,1\}^n$, $y\in \{0,1\}^{4n}$ such that $u_{x,y}=v_{x,y}$. For a random function $f:\{0,1\}^n\times\{0,1\}^{4n}\to \{0,1\}^{12n}$, the probability that there exists a pair $x,y$ such that $u_{x,y}=v_{x,y}$ is $\leq \frac{2^{5n}}{2^{6n}} = 2^{-n}$ by the union bound. But since
    $$\sum_{n=1}^{\infty} 2^{-n} < \infty$$
    by Borel-Cantelli the probability this happens infinitely often is $0$. Thus, with probability $1$ over $\O\gets \D$, correctness is satisfied for all sufficiently large $n$.
\end{proof}

\begin{claim}
    With probability $1$ over $\O\gets \D$,~\Cref{cons:claw} satisfies search security.
\end{claim}

\begin{proof}
    Fix some polynomial $p(n),q(n)$. We can assume without loss of generality that any adversary breaking search security outputs $(y,u_{x,y},v_{x,y})$ instead of $(y,v_{x,y},u_{x,y})$. This loses at most a factor of $2$ in the security by randomly reordering the strings.
    
    Let $BAD_{n,p,q}^{(0)}$ be the event (over randomness the choice of oracle) that for some circuit $C^{\O,\PSPACE}$ of size $\leq p(n)$ we have
    $$\Pr_{x\gets \{0,1\}^n}[C^{\O,\PSPACE}(x) = (y,u_{x,y},v_{x,y})\text{ for some }y\in \{0,1\}^{4n}] \geq \frac{1}{q(n)}$$

    To prove the theorem, it suffices to show that for all polynomials $p,q$, 
    $$\Pr_{\O\gets \D}[BAD_{n,p,q}^{(0)}\text{ occurs infinitely often}] = 0$$
    We will first show that $\Pr[BAD_{n,p,q}^{(0)}] = 2^{-\Omega(n)}$.

    Let $\wt{p}(n)$ be a polynomial upper bound on the size of the entire game. Let $\Sim_{\aux}$ and $L(n)$ be as in~\Cref{thm:keylemma} with accuracy parameter $8q(n)$ and circuit-size bound $\wt{p}(n)$. Define a new event $BAD_{n,p,q}^{(1)}$ to be the event that for some circuit $C^{\O,\PSPACE}$ of size $\leq p(n)$, we have
    $$\Pr_{x\gets \{0,1\}^n,\wt{\RO}_n}[C^{\Sim_{\aux}(\aux_{\wt{\RO}_n}),\PSPACE}(x) = (y,\wt{\RO}_n(x,y))\text{ for some }y] \geq \frac{1}{2q(n)}$$
    Applying~\Cref{cor:truerand} with error $\frac{1}{4q(n)}$ and circuit-size bound $\wt{p}(n)$ to the entire game (including the sampling of $x$ using $\SG_n$ and final computation of $\RO_n(x,y)$ on input $(x,y)$ using $\ket{\phi_x}$) gives that if $BAD_{n,p,q}^{(0)}$ holds and the event in ~\Cref{cor:truerand} holds, we have that
    \begin{equation}\label{eq:switchrand}
        \Pr_{x\gets \{0,1\}^n,\wt{\SG}_n,\wt{\Expand}_n,\wt{\Mix}_n}\left[C^{\wt{\SG}_n,\wt{\Expand}_n,\wt{\Mix}_n,\PSPACE}(x) = (y,\wt{\RO}_n(x,y))\text{ for some }y\right] \geq \frac{3}{4q(n)}
    \end{equation}
    We then use the hybrids in the proof of~\Cref{thm:keylemma}, keeping $\wt{\RO}_n$ fixed and testing whether $C$ outputs $(y,\wt{\RO}_n(x,y))$. Since $C$ is independent of the fresh random unitaries and states, $\Hyb_0$ and $\Hyb_1$ have the same distribution after averaging. The remaining hybrids change the probability of success by at most $1/(8q(n))$, since the final test makes no oracle queries. In particular, if~\Cref{eq:switchrand} holds then
    $$\Pr_{x\gets \{0,1\}^n,\wt{\RO}_n}[C^{\Sim_{\aux}(\aux_{\wt{\RO}_n}),\PSPACE}(x) = (y,\wt{\RO}_n(x,y))\text{ for some }y] \geq \frac{3}{4q(n)} - \frac{1}{8q(n)} \geq \frac{1}{2q(n)}$$
    and so
    $BAD_{n,p,q}^{(1)}$ also holds. And so by the union bound,
    $$\Pr[BAD_{n,p,q}^{(0)}] \leq \Pr[BAD_{n,p,q}^{(1)}] + 2^{-\Omega(n)}$$

    Let $c(\O)$ describe a circuit witnessing $BAD_{n,p,q}^{(1)}$ when one exists, and any circuit otherwise. Below, $\Eval(c(\O),x)$ runs the described circuit on $x$. Write $$p_{BAD} = \Pr_{\O\gets \D,x,\wt{\RO}_n}[\Eval^{\Sim_{\aux}(\aux_{\wt{\RO}_n}),\PSPACE}(c(\O),x) = (y,\wt{\RO}_n(x,y))\text{ for some }y]$$

    Observe that we can write the left hand side in terms of $\Pr[BAD_{n,p,q}^{(1)}]$ as follows:
    \begin{equation*}
        \begin{split}
            &\Pr_{\O\gets \D,x,\wt{\RO}_n}[\Eval^{\Sim_{\aux}(\aux_{\wt{\RO}_n}),\PSPACE}(c(\O),x) = (y,\wt{\RO}_n(x,y))\text{ for some }y]\\
            \geq& \Pr_{\O\gets \D,x,\wt{\RO}_n}[\Eval^{\Sim_{\aux}(\aux_{\wt{\RO}_n}),\PSPACE}(c(\O),x) = (y,\wt{\RO}_n(x,y))\text{ for some }y|BAD_{n,p,q}^{(1)}]\\
            &\qquad{}\cdot \Pr[BAD_{n,p,q}^{(1)}]\\
            \geq& \frac{\Pr[BAD_{n,p,q}^{(1)}]}{2q(n)}
        \end{split}
    \end{equation*}

    And so it suffices to bound $p_{BAD}$. Write $L=L(n)$ and consider $J_x = (y_{x,1},\dots,y_{x,L})$. Let $DIST_x$ be the event that these are all distinct. First, $\Pr[x\in K^\ast]\leq L2^{-n}$ and $\Pr[\neg DIST_x]\leq L^2 2^{-4n}$. Condition on neither event occurring, and fix $x$, $K$, all sampled labels, and $c(\O)$.

    Since $\wt{\RO}_n$ is fresh and independent of $\O$, the pairs $\wt{\RO}_n(x,y_{x,i})$ remain independent and uniform on $\{0,1\}^{12n}$. The only advice depending on $\wt{RO}_n(x,\cdot)$ consists of the corresponding $L$ claw states. We may therefore view the entire simulated computation as a decoder receiving these claw states and independent auxiliary input. Applying~\Cref{lem:claw-hardness} with $m=6n$ and $\ell=L$, and discarding the decoder's output label $y$, bounds the probability of recovering a pair at one of the sampled points by $L2^{-6n}$.

    At an unsampled point, $\wt{\RO}_n(x,y)$ is independent of all the advice, so guessing it correctly has probability at most $2^{-12n}$, even when $y$ is chosen as a function of the simulated computation. Averaging over the fixed data, we obtain
    $$p_{BAD}\leq L2^{-n}+L^2 2^{-4n}+L2^{-6n}+2^{-12n}\leq 2L2^{-n}+L^2 2^{-4n}.$$

    Thus,
    $$\Pr[BAD_{n,p,q}^{(1)}] \leq 6q(n)\cdot L\cdot 2^{-n} \leq 2^{-n/2}$$

    And so with probability $1-2^{-\Omega(n)}$, $BAD_{n,p,q}^{(0)}$ does not occur.

    Thus, for all polynomials $p,q$, 
    $$\sum_{n=1}^{\infty} \Pr[BAD^{(0)}_{n,p,q}] < \infty$$
    and so by Borel-Cantelli, for all $p,q$,
    $$\Pr_{\O\gets \D}[BAD^{(0)}_{n,p,q}\text{ occurs infinitely often}]=0$$
\end{proof}

By~\Cref{thm:csg-implies-osp} and the remark following it, we obtain qv-OSP relative to $\O,\PSPACE$.

\section{AI Disclosure}

The proofs of all theorems were developed in discussion with various versions of ChatGPT. The description of the oracles and most of the key ideas for the proofs were human originated. The primary exception to this is the technique for removing the Forrelated oracle in the non-uniform setting by considering the Turing machine whose success probability approximates the probability of the existence of a non-uniform attacker. This corresponds to the first half of the proof of~\Cref{thm:weakhard}. LLMs were also used to generate the proofs in~\Cref{sec:commitment-ev-puzzles}.

\bibliographystyle{alpha}
\bibliography{refs/abbrev0,refs/crypto,refs/refs}

\clearpage
\appendix
\section{From state preparation channels to unitaries}
\label{sec:stateprepcollapse}

Let $\vec{\sigma}=\{\sigma_t\}_{t\in\N}$ be any fixed family of states, where $\sigma_t$ is a state on $t$ qubits. Fix a purification of each $\sigma_t$ on $2t$ qubits, and independently sample $U_t\gets\Haar(\{0,1\}^t)$. Write $S=\{S_{\sigma_t}^{U_t}\}_{t\in\N}$. Access to $\vec{\sigma}$ means access to a channel which, on input $t$, returns a fresh copy of $\sigma_t$. Queries to level $t$ take at least $t$ time, and we allow controlled queries to $S$.

Throughout this section, complexity classes refer to promise problems. An oracle for $\QCMA^{\vec{\sigma}}$ means an oracle for its complete verification promise problem: given a verifier and a runtime bound in unary, distinguish whether some classical witness is accepted with probability at least $2/3$, or every classical witness is accepted with probability at most $1/3$. Its behavior outside the promise is arbitrary.

\begin{theorem}\label{thm:stateprepcollapse}
    With probability $1$ over $\{U_t\}_{t\in\N}$,
    $$\QCMA^S\subseteq\BQP^{S,\QCMA^{\vec{\sigma}}}.$$
\end{theorem}

\begin{proof}
    Let $\Pi$ be any promise problem in $\QCMA^S$ with verifier $\Ver^S$. By amplification, assume that its completeness is at least $11/12$ and its soundness is at most $1/12$. Let $q(n)\geq n$ be a polynomial upper bound on its runtime, witness length, and circuit size after hardcoding $x,w$. In particular, it makes at most $q(n)$ queries, all to levels at most $q(n)$. We describe a QPT algorithm $\A^{S,\QCMA^{\vec{\sigma}}}$ as follows:
    \begin{enumerate}
        \item On input $x\in\{0,1\}^n$, let $d=\ceil{c\log n}$ for a sufficiently large constant $c$.
        \item For each $t\in[1,d]$, apply~\Cref{thm:tomography} to $S_{\sigma_t}^{U_t}$ with error parameter $1/q(n)^3$ and failure probability $2^{-d}$. This takes $2^{O(d)}\poly(n)=\poly(n)$ time. Let $\wt S_t$ be the resulting circuit.
        \item Construct a new verifier $\wt\Ver^{\vec{\sigma}}$ which hardcodes $x$ and the descriptions of the circuits $\wt S_t$. It runs $\Ver^S(x,w)$, replacing queries to levels $t\leq d$ by $\wt S_t$. For each $t>d$, replace queries by the stateful simulator from~\Cref{lem:stateprep}, with query bound $q(n)$ and error $1/q(n)^2$, using fresh copies of $\sigma_t$. Use a separate simulator for each level, retaining its state between queries.
        \item Query the complete $\QCMA^{\vec{\sigma}}$ verification promise problem on $\wt\Ver^{\vec{\sigma}}$ and output its answer.
    \end{enumerate}
    Each simulator uses $O(q(n)^4)$ copies, and there are at most $q(n)$ levels. Thus, $\wt\Ver^{\vec{\sigma}}$ has polynomial runtime and its description has polynomial length.

    To show correctness, first leave the oracles at levels at most $d$ unchanged and average over independent Haar unitaries at the larger levels. For $\ell\in[d,q(n)]$, define
    $$a_\ell(x,w)=\E_{\{U'_t\}_{t>\ell}}\left[\Pr\left[\Ver^{S_{\sigma_1}^{U_1},\dots,S_{\sigma_\ell}^{U_\ell},S_{\sigma_{\ell+1}}^{U'_{\ell+1}},\dots,S_{\sigma_{q(n)}}^{U'_{q(n)}}}(x,w)=1\right]\right].$$
    In particular, $a_{q(n)}(x,w)=\Pr[\Ver^S(x,w)=1]$. For fixed $U_1,\dots,U_{\ell-1}$, the expectation of $a_\ell(x,w)$ over $U_\ell$ is $a_{\ell-1}(x,w)$. Each query to $S_{\sigma_\ell}^{U_\ell}=C_{U_\ell}S_{\sigma_\ell}C_{U_\ell}^\dagger$ uses at most two queries to $U_\ell$ or its inverse. Thus, by~\Cref{thm:unitarysim},
    $$\Pr\left[\abs{a_\ell(x,w)-a_{\ell-1}(x,w)}>\frac{1}{q(n)^2}\right]
    \leq 2\exp\left(-\frac{2^\ell-2}{384q(n)^6}\right).$$
    The other oracle levels can be treated as fixed side information or independent randomness in this application. Taking a union bound over $\ell>d$, $x\in\{0,1\}^n$, and $w\in\{0,1\}^{q(n)}$, the probability that any of these bounds fails is at most
    $$2^{n+q(n)+1}q(n)\exp\left(-\frac{2^{d+1}-2}{384q(n)^6}\right)\leq\frac{1}{n^2}$$
    for all sufficiently large $n$, by choosing $c$ sufficiently large. Whenever none fails, simultaneously for all $x,w$,
    $$\abs{\Pr[\Ver^S(x,w)=1]-a_d(x,w)}\leq\frac{1}{q(n)}.$$

    By~\Cref{lem:stateprep}, replacing the averaged oracles at levels above $d$ by their simulators changes acceptance probabilities by at most $q(n)/q(n)^2=1/q(n)$. This follows by replacing one level at a time, keeping the other levels as side oracles. Whenever all tomography estimates are accurate, replacing the remaining oracles by $\wt S_t$ changes acceptance probabilities by at most $q(n)/q(n)^3=1/q(n)^2$. These bounds hold for all $x,w$, so
    $$\abs{\Pr[\Ver^S(x,w)=1]-\Pr[\wt\Ver^{\vec{\sigma}}(w)=1]}
    \leq\frac{2}{q(n)}+\frac{1}{q(n)^2}\leq\frac{1}{12}$$
    for all sufficiently large $n$. Consequently, on a yes input some witness is accepted by $\wt\Ver^{\vec{\sigma}}$ with probability at least $5/6$, and on a no input every witness is accepted with probability at most $1/6$. The query to the $\QCMA^{\vec{\sigma}}$ oracle is therefore within its promise and gives the correct answer.

    Except for a set of choices of $\{U_t\}_t$ of probability at most $1/n^2$, the algorithm is correct on every promised input of length $n$ whenever tomography succeeds. Tomography succeeds with probability at least $1-d2^{-d}\geq 5/6$ for all sufficiently large $n$. By Borel--Cantelli, with probability $1$ over $\{U_t\}_t$, the algorithm correctly decides $\Pi$ on all sufficiently large input lengths. Hardcoding the answers at the finitely many remaining lengths gives an algorithm deciding $\Pi$ on all promised inputs. Finally, there are only countably many verifiers, so with probability $1$ the containment holds for all of them simultaneously.
\end{proof}

\begin{corollary}
    If $\BQP^{\vec{\sigma}}=\QCMA^{\vec{\sigma}}$, then with probability $1$ over $\{U_t\}_{t\in\N}$,
    $$\BQP^S=\QCMA^S.$$
\end{corollary}

\begin{proof}
    By~\Cref{thm:stateprepcollapse}, it suffices to implement the $\QCMA^{\vec{\sigma}}$ verification oracle in $\BQP^S$. By assumption, this promise problem has a $\BQP^{\vec{\sigma}}$ algorithm, which we amplify to have error at most $1/12$. Each copy of $\sigma_t$ needed by this algorithm can be prepared by applying $S_{\sigma_t}^{U_t}$ to the all-zero state and tracing out the purification register. Thus, $\QCMA^S\subseteq\BQP^S$, and the reverse containment is immediate.
\end{proof}

\section{EV-OWPuzzs from OSP}
\label{sec:commitment-ev-puzzles}

\subsection{Commitments and the implication from OSP}

\begin{definition}[Commitments with a classical committer~\cite{C:BarKhu25}]
\label{def:classical-committer-commitment}
A statistically sum-binding, computationally hiding commitment with a classical committer and a quantum receiver consists of four algorithms $(\mathsf{Com},\mathsf{Rec},\mathsf{Open},\Ver)$ with the following syntax.
In the commit phase, the parties interact as follows:
\begin{align}
(\mathsf{st},\rho)\leftarrow
\langle\mathsf{Com}(1^n,b),\mathsf{Rec}(1^n)\rangle,
\end{align}
where $\mathsf{Com}$ is PPT, $\mathsf{Rec}$ is QPT, and all communication is classical.
The committer's output $\mathsf{st}$ is classical, whereas the receiver's output $\rho$ may be quantum.
In the opening phase, the PPT algorithm $\mathsf{Open}(\mathsf{st})$ outputs a single classical message $(b,d)$, where $b$ is the original input bit.
The QPT algorithm $\Ver(\rho,b,d)$ accepts or rejects.
We require the following properties.
\begin{itemize}
\item \emph{Correctness.}
For each $b\in\{0,1\}$, an honest opening to $b$ is accepted with probability at least $1-\negl(n)$.
\item \emph{Computational hiding.}
The final views of every nonuniform QPT receiver interacting with $\mathsf{Com}(1^n,0)$ and $\mathsf{Com}(1^n,1)$ are computationally indistinguishable.
\item \emph{Statistical sum-binding.}
Consider any computationally unbounded adversarial committer $\mathcal{C}$ that interacts with the honest receiver in the commit phase and retains a private state $\sigma$.
After this phase, sample an independent uniform bit $b'$, and let $\mathcal{C}$, given $\sigma$ and $b'$, output an opening string $d$.
For every such $\mathcal{C}$, there is a negligible function $\eta_{\mathcal{C}}$ such that
\begin{align}
\Pr[\Ver(\rho,b',d)=1]\le\frac12+\eta_{\mathcal{C}}(n).
\label{eq:commitment-ev-binding}
\end{align}
If the committer does not provide an opening string, the receiver rejects.
\end{itemize}
\end{definition}

\begin{theorem}[Bartusek--Khurana~\cite{C:BarKhu25}]
\label{thm:osp-classical-commitment}
If OSP exist, then commitments as in \Cref{def:classical-committer-commitment} exist.
\end{theorem}

\begin{theorem}\label{thm:commitment-ev-puzzles}
If commitments as in \Cref{def:classical-committer-commitment} exist, then EV-OWPuzzs exist with perfect correctness, a QPT sampler, and a classical deterministic polynomial-time verifier.
\end{theorem}

\subsection{The transcript determines the committed bit}

Run the honest commit phase with a uniform bit $B$.
Let $R$ be the committer's random tape for this phase, padded to a fixed polynomial length, and let $T$ be the transcript.
Write $W=(B,R)\in\{0,1\}^m$, where $m=m(n)=\poly(n)$, so that $W$ is uniform.
For $w=(b,r)$, define $\mathsf{Cons}(t,w)=1$ if replaying $\mathsf{Com}(1^n,b)$ with tape $r$ and the receiver messages in $t$ reproduces all the committer messages in $t$ and terminates after exactly the messages in $t$.
Otherwise set $\mathsf{Cons}(t,w)=0$.
This is a classical deterministic polynomial-time test, including on malformed transcripts.
For each fixed bit $b\in\{0,1\}$, define
\begin{align}
S_t&\coloneqq\{w\in\{0,1\}^m:\mathsf{Cons}(t,w)=1\},\qquad
S_{t,b}\coloneqq\{(\beta,r)\in S_t:\beta=b\},\notag\\
p_t&\coloneqq\Pr[B=0\mid T=t].
\end{align}
All conditional statements below concern transcripts $t$ with $\Pr[T=t]>0$.

\begin{lemma}\label{lem:commitment-ev-conditioning}
Conditioned on $T=t$, the tape $W$ is uniform over $S_t$.
Moreover, the receiver's conditional state $\rho_t$ is independent of $W$.
In particular, $p_t=|S_{t,0}|/|S_t|$.
\end{lemma}

\begin{proof}
For a fixed tape $w\in S_t$, the probability of the transcript $t$ is the probability that the receiver sends the receiver messages in $t$ upon receiving the committer messages in $t$.
This probability depends only on $t$, and is zero for $w\notin S_t$.
Since $W$ is uniform, Bayes' rule proves the first assertion.
The receiver starts in a state independent of $W$ and communicates with the committer only classically.
Conditioned on its incoming and outgoing messages, its final state is therefore determined by $t$ alone, proving the second assertion.
\end{proof}

\begin{lemma}\label{lem:commitment-ev-concentration}
There is a negligible function $\nu$ such that
\begin{align}
\E_T[\min\{p_T,1-p_T\}]\le\nu(n).
\label{eq:commitment-ev-concentration}
\end{align}
Consequently, $\E_T[p_T(1-p_T)]\le\nu(n)$.
\end{lemma}

\begin{proof}
For $b\in\{0,1\}$ with $S_{t,b}\ne\varnothing$, define $q_b(t)$ by the following experiment.
Sample $w=(b,r)$ uniformly from $S_{t,b}$, replay the committer to reconstruct its final state $\mathsf{st}$, and run $\mathsf{Open}(\mathsf{st})$ with fresh randomness.
Let $q_b(t)$ be the probability that this opening to $b$ is accepted against $\rho_t$.
Set $q_b(t)=0$ when $S_{t,b}=\varnothing$.
By \Cref{lem:commitment-ev-conditioning}, this experiment gives exactly the honest opening success probability conditioned on $T=t$ and $B=b$, whenever this event has positive probability.
Thus correctness gives
\begin{align}
\E_T[p_Tq_0(T)+(1-p_T)q_1(T)]\ge1-\negl(n).
\label{eq:commitment-ev-honest-open}
\end{align}

Consider the following unbounded committer $\mathcal{C}$ in the sum-binding experiment.
In the commit phase, it samples a uniform bit and follows the honest committer, retaining $t$.
Given the target bit $b'=b$ in the opening phase, it enumerates $S_{t,b}$.
If this set is empty, it aborts; otherwise, it samples a uniform tape from this set, reconstructs $\mathsf{st}$, and runs $\mathsf{Open}(\mathsf{st})$.
Its commit phase has the honest transcript distribution, and the receiver's conditional state is $\rho_t$ regardless of the tape originally used.
Its opening success probability is therefore $\E_T[(q_0(T)+q_1(T))/2]$.
By \cref{eq:commitment-ev-binding}, for a negligible $\eta=\eta_{\mathcal{C}}$,
\begin{align}
\E_T\left[\frac{q_0(T)+q_1(T)}2\right]\le\frac12+\eta(n).
\label{eq:commitment-ev-resampled-open}
\end{align}

For every $p,q_0,q_1\in[0,1]$,
\begin{align}
pq_0+(1-p)q_1
&=\frac{q_0+q_1}{2}+\left(p-\frac12\right)(q_0-q_1)\notag\\
&\le\frac{q_0+q_1}{2}+\left|p-\frac12\right|.
\end{align}
Combining this inequality with \cref{eq:commitment-ev-honest-open,eq:commitment-ev-resampled-open} yields
\begin{align}
\E_T[\min\{p_T,1-p_T\}]
=\frac12-\E_T\left[\left|p_T-\frac12\right|\right]
\le\negl(n)+\eta(n)=\negl(n).
\end{align}
The second claim follows from $p(1-p)\le\min\{p,1-p\}$.
\end{proof}

\subsection{Construction and proof of security}

Let $L=2^{\lceil\log_2(m+3)\rceil}$.
For $0\le\ell<L$, let $\mathcal{H}_\ell=\{h_A:A\in\{0,1\}^{\ell\times m}\}$, where $h_A(w)=Aw$ over $\mathbb{F}_2$.
These families can be sampled and evaluated in classical polynomial time; $\mathcal{H}_0$ consists of the map with empty output.
For distinct $w,w'$, the vector $A(w-w')$ is uniform over $\{0,1\}^\ell$ when $A$ is uniform, so
\begin{align}
\Pr_{h\leftarrow\mathcal{H}_\ell}[h(w)=h(w')]=2^{-\ell}
\quad\text{for all distinct }w,w'\in\{0,1\}^m.
\end{align}
The sampler $\Samp_0(1^n)$ does the following.
\begin{enumerate}
\item Sample $W=(B,R)\leftarrow\{0,1\}^m$ and run the honest commit phase with input $B$ and committer tape $R$, obtaining the transcript $T$.
\item Independently sample $\ell\leftarrow\{0,\ldots,L-1\}$ and $h\leftarrow\mathcal{H}_\ell$ uniformly.
\item Output the puzzle $x=(1^n,T,\ell,h,h(W))$ and the answer $W$.
\end{enumerate}
On input $x=(1^n,t,\ell,h,z)$ and an answer $w'$, the verifier $\Ver_0$ accepts exactly when
\begin{align}
\mathsf{Cons}(t,w')=1\quad\text{and}\quad h(w')=z,
\label{eq:commitment-ev-verify}
\end{align}
Malformed inputs are rejected.
The sampler is QPT and the verifier is deterministic classical polynomial-time.
Correctness is perfect because the actual tape always reproduces the actual transcript and passes the hash check.

\begin{lemma}[Weak security]\label{lem:commitment-ev-weaksecurity}
For every nonuniform QPT algorithm $\A$ and all sufficiently large $n$,
\begin{align}
\Pr[\Ver_0(x,\A(x))=1:(x,w)\leftarrow\Samp_0(1^n)]
\le1-\frac{1}{64L(n)}.
\label{eq:commitment-ev-weaksecurity}
\end{align}
\end{lemma}

\begin{proof}
Suppose a nonuniform QPT solver $\A$ violates \cref{eq:commitment-ev-weaksecurity} for infinitely many $n$, and fix such an $n$.
Construct a nonuniform QPT algorithm $\mathcal{F}$ as follows.
Given $1^n,t$, for each $\ell=L-1,L-2,\ldots,0$, run $n$ independent trials in which $h\leftarrow\mathcal{H}_\ell$ and $z\leftarrow\{0,1\}^\ell$ are sampled uniformly and $\A$ is run on $(1^n,t,\ell,h,z)$.
Return the first answer accepted by $\Ver_0$, or $\bot$ if none is accepted.
Each trial uses a fresh execution of $\A$.
Whenever $\mathcal{F}$ returns a value other than $\bot$, that value lies in $S_t$.

For each possible honest transcript $t$, set
\begin{align}
k_t\coloneqq\lceil\log_2|S_t|\rceil+2.
\end{align}
Since $1\le|S_t|\le2^m$, this length occurs in the search and satisfies
\begin{align}
k_t\le L-1,\qquad 4|S_t|\le2^{k_t}<8|S_t|.
\label{eq:commitment-ev-critical}
\end{align}
Let $e_t$ be the failure probability of $\A$ on a genuine puzzle conditioned on $T=t$ and $\ell=k_t$.
By \Cref{lem:commitment-ev-conditioning}, such a puzzle uses a uniform $w\leftarrow S_t$ and target $h(w)$.
Since the sampler chooses each length with probability $1/L$, the assumed solver failure probability gives
\begin{align}
\E_T[e_T]<\frac1{64}.
\end{align}
Call $t$ good if $e_t\le1/4$.
By Markov's inequality, $\Pr[T\text{ is not good}]\le1/16$.

\paragraph{Success at the critical length.}
Fix a good transcript $t$.
For any fixed $w\in S_t$, a random $h\leftarrow\mathcal{H}_{k_t}$ maps another tape in $S_t$ to $h(w)$ with probability at most $(|S_t|-1)2^{-k_t}\le1/4$, by the union bound.
Hence, on a genuine puzzle, $\A$ succeeds and $w$ has a unique hash value with probability at least $1-e_t-1/4\ge1/2$.
For each fixed $h$, these tapes give distinct targets $h(w)$.
Each such target has probability $1/|S_t|$ when sampling $w\leftarrow S_t$ and using $h(w)$, and probability $2^{-k_t}$ when sampling $z$ uniformly.
Restricting to these targets and averaging over $h$ gives
\begin{align}
\Pr[\text{a trial at length }k_t\text{ succeeds}]
&\ge\frac{|S_t|}{2^{k_t}}\left(1-e_t-\frac14\right)\notag\\
&\ge\frac18\cdot\frac12=\frac1{16}.
\label{eq:commitment-ev-trial}
\end{align}
Thus all $n$ independent trials at this length fail with probability at most $(15/16)^n$.

\paragraph{Excluding tapes with the wrong bit.}
Condition on $T=t$ and $B=b$.
To accept a tape with bit $1-b$, a trial must choose $z\in h(S_{t,1-b})$.
For each fixed $h$, this set has at most $|S_{t,1-b}|$ elements, so a uniform length-$\ell$ target hits it with probability at most $|S_{t,1-b}|2^{-\ell}$.
For the analysis, run all trials, including those $\mathcal{F}$ would skip after returning.
Let $\mathsf{Wrong}$ denote the event that some trial with $\ell\ge k_t$ accepts a tape with bit $1-b$.
The union bound over the $n$ trials at each length gives
\begin{align}
\Pr[\mathsf{Wrong}\mid T=t,B=b]
&\le n|S_{t,1-b}|\sum_{\ell=k_t}^{L-1}2^{-\ell}\notag\\
&\le\frac{n|S_{t,1-b}|}{2|S_t|},
\label{eq:commitment-ev-wrong-bucket}
\end{align}
since $\sum_{\ell=k_t}^{L-1}2^{-\ell}\le2^{1-k_t}$ and $2^{k_t}\ge4|S_t|$.
By \Cref{lem:commitment-ev-conditioning}, averaging over $B$ gives
$\Pr[\mathsf{Wrong}\mid T=t]\allowbreak\le\frac n2[p_t(1-p_t)+(1-p_t)p_t]=np_t(1-p_t)$.

For good $t$, some trial at length $k_t$ succeeds except with probability $(15/16)^n$.
Since $\mathcal{F}$ searches lengths in decreasing order, it then returns at length at least $k_t$.
Unless $\mathsf{Wrong}$ occurs, this output has bit $B$.
Combining these bounds with $\Pr[T\text{ is not good}]\le1/16$ and \Cref{lem:commitment-ev-concentration} gives
\begin{align}
\Pr[\mathcal{F}(1^n,T)\text{ returns a tape whose bit is }B]
&\ge1-\frac1{16}-\left(\frac{15}{16}\right)^n-n\E_T[p_T(1-p_T)]\notag\\
&\ge\frac{15}{16}-\negl(n).
\end{align}
A nonuniform QPT receiver can follow the honest commit phase and run $\mathcal{F}(1^n,T)$, outputting the bit of the returned tape or a uniform bit if $\mathcal{F}$ returns $\bot$.
This contradicts computational hiding for infinitely many $n$.
\end{proof}

\begin{proof}[Proof of \Cref{thm:commitment-ev-puzzles}]
The construction above has perfect correctness, a QPT sampler, and a classical deterministic polynomial-time verifier.
By \Cref{lem:commitment-ev-weaksecurity}, its security gap $1/(64L(n))$ is inverse polynomial.
Apply the security amplification theorem for EV-OWPuzzs of Chung, Goldin, and Gray~\cite{C:ChuGolGra24}.
The resulting puzzle consists of polynomially many independent component puzzles and accepts exactly when every component answer is accepted.
Its solving probability is negligible, and its correctness remains perfect.
The sampler remains QPT and the verifier remains classical, deterministic, and polynomial-time.
\end{proof}

\end{document}